\documentclass[10pt,reqno,final]{article}

\usepackage{amsmath,amsfonts,amssymb,amsthm,version}
\usepackage{mathrsfs,fancybox,pifont}
\usepackage{graphicx}
\usepackage{url,hyperref}
\usepackage[notcite,notref]{showkeys}
\usepackage{xcolor}
\usepackage{subfigure,multirow}
\usepackage{epstopdf}
\usepackage{cases}
\usepackage[shortlabels]{enumitem}
\usepackage{mathtools}
\usepackage{algorithm,algorithmic}
\usepackage{authblk}
\usepackage{lipsum}
\usepackage{MnSymbol}

\allowdisplaybreaks

\numberwithin{equation}{section}
\numberwithin{figure}{section}
\numberwithin{table}{section}

\theoremstyle{plain}
\newtheorem{theorem}{Theorem}[section]
\newtheorem{lemma}[theorem]{Lemma}

\theoremstyle{definition}

\theoremstyle{remark}

\title{A FORMULA FOR CONNECTED BOSONIC $N$-POINT FUNCTIONS FOR BKP HIERARCHY}
\author[1]{XUHUI ZHANG\thanks{zhangxh19@mails.tsinghua.edu.cn}}
\author[1]{JIAN ZHOU\thanks{jianzhou@mail.tsinghua.edu.cn}}
\affil[1]{Department of Mathematical Science, Tsinghua University}
\date{\today}

\begin{document}
\maketitle

\begin{abstract}
    We present a formula for the connected \(n\)-point functions of a tau-funtion of the BKP hierarchy by embedding BKP hierarchy into KP hierarchy. This formula is different from the one given by Wang and Yang \cite{wangBKPHierarchyAffine2022}. We prove that these two formulae are equivalent.

    \medskip
    \noindent{\bf Keywords}: KP hierarchy, BKP hierarchy, connected bosoinc \(n\)-piont function
\end{abstract}

% \tableofcontents

\newpage

\section{Introduction}
It is an important problem to find tau-functions and their corresponding (connected) \(n\)-point functions in mathematical physics.
For the topological 2D gravity theory whose partition function is given by the
Witten-Kontsevich tau-function \cite{wittenTwodimensionalGravityIntersection1990, kontsevichIntersectionTheoryModuli1992},
the one-point function, two-function, and the three-function are found by Witten, Dijkgraaf and Zagier
respectively.
Okounkov \cite{okounkovGeneratingFunctionsIntersection2002} expressed the \(n\)-point functions in this case as $n$-dimensional error-function-type integrals and gave a new derivation of Witten's KdV equations.
In \cite{liuNpointFunctionsIntersection2009}, Liu and Xu found a recursion formula for the \(n\)-point functions, and they used it to prove the Faber conjecture, see \cite{liuNewPropertiesIntersection2007} and \cite{liuProofFaberIntersection2009}.
Buryak {\em et al}. proved a new formula in \cite{buryakINTEGRALSPsiCLASSES} and Alexandrov et al. \cite{alexandrov2021buryak} gave yet another derivation of
the Witten Conjecture/Kontsevich Theorem.
It involves a summation over all permutations $\tau$ of $\{1, \dots, n\}$ such that $\tau(1) = 1$
and as in the Okounkov formula, an integral over $n$-dimensional space.

More recently, a different type of formulae for $n$-point functions for various theories
governed by integrable hierarchies
for tau-functions have been derived using matrix resolvent.
In \cite{bertolaCorrelationFunctionsKdV2016}, Bertola, Dubrovin and Yang found an explicit connected \(n\)-point functions of an arbitrary tau-function of the KdV hierarchy in terms of matrix resolvents
and a summation over permutations.
Dubrovin-Yang \cite{dubrovinGeneratingSeriesGUE2017} used the similar method to find the connected \(n\)-point functions of tau-functions of Toda lattice hierarchy in terms of matrix resolvents of the difference Lax operator, and as an application, they obtained explicit generating series for connected GUE correlators. See \cite{dubrovinTaufunctionsKdVHierarchy2021} and \cite{yangTaufunctionsTodaLattice2021} for another proof and some new formulae. Moreover, Mattia and Yang \cite{cafassoTaufunctionsAblowitzLadik2021} extended the matrix-resolvent method for computing the connected \(n\)-point functions of tau-functions of the Ablowitz-Ladik hierarchy.

Inspired by the above work,
yet another type of formulae for $n$-point function has been derived.
In \cite{zhouExplicitFormulaWittenKontsevich2013}, the second author found an explicit formula 
for the fermionic representation of Witten-Kontsevich tau-function.
 Later, Balogh and Yang \cite{baloghGeometricInterpretationZhou2017} proved that the coefficients given in \cite{zhouExplicitFormulaWittenKontsevich2013} are the affine coordinates for the point of the Sato Grassmannian corresponding to the Witten-Kontsevich tau function.
Combining the above ideas,
Zhou \cite{zhouEmergentGeometryMirror2015} gave a formula for the connected bosoinc \(n\)-point function of KP hierarchy directly in terms of the affine coordinates.
In particular a new formula for the $n$-point functions of the Witten-Kontsevich tau-function
in terms of Airy functions is obtained by this approach.

The key ingredient in deriving the formula in \cite{zhouEmergentGeometryMirror2015} is
the boson-fermion correspondence applied to the study of KP hierarchy.
This has played a crucial role in Kyoto school approach to integrable hierarchies \cite{miwaSolitonsDifferentialEquations2000}.
They emphasized on the symmetries of the integrable hierarchies,
and this leads naturally to infinite-dimensional Lie algebras and representations \cite{jimboSolitonsInfiniteDimensional1983, kacModularConformalInvariance1988, drinfeldLieAlgebrasEquations}.
In particular, the
KP hierarchy corresponds to the basic representation of the Lie algebra of type $A_\infty$,
and the BKP hierarchy corresponds to the spin representation of the Lie algebra of type $B_\infty$.
In \cite{jimboSolitonsInfiniteDimensional1983} various reductions of the KP hierarchy have been considered by
considering the fermionic constructions of the representations of the infinite-dimensional Lie algebras.
Such constructions are obtained by taking the invariant subspaces of suitable involutions
on the fermionic Fock space.
This has inspired Zhou \cite{zhouFermionicComputationsIntegrable2015} to propose the approach to various integrable hierarchies
by embedding them into the KP hierarchy.
This can be achieved by suitable embedding of the Lie algebras of their transformation groups into the Lie algebra of type $A_\infty$.
On the other hand,
for the Lie algebras in \cite{jimboSolitonsInfiniteDimensional1983},
their corresponding integrable hierarchies can also be studied by
applying the corresponding boson-fermion correspondences.
A priori,
the above two approaches should lead to equivalent results,
however the appearance of the results may be different,
and to establish  their equivalence can be a challenging problem.
In this paper,
we will address such issues for the formulae for $n$-point functions of the BKP hierarchy.

Using the boson-fermion correspondence of type B,
Wang and Yang \cite{wangBKPHierarchyAffine2022} modified Zhou's method to get a formula for the connected bosoinc \(n\)-point function of tau-function of BKP hierarchy. More recently they generalize to the cases of
connected $(n,m)$-point functions of 2D Toda lattice hierarchy \cite{wang2023diagonal}
and diagonal 2-BKP hierarchy \cite{wang2023connected}.
In this paper we first apply the approach proposed in  \cite{zhouFermionicComputationsIntegrable2015} to derive another formula for the BKP case from the formula for the KP hierarchy,
and then we prove that our formula is equivalent to the formula in \cite{wangBKPHierarchyAffine2022}.
Our method for deriving the formula for $n$-point functions is clearly applicable to other reductions of the KP hierarchy.
It has the advantage of not depending on the specific boson-fermion correspondence
for the reduction.

Given a \(\tau\)-function \(\tau^{BKP}\) of the BKP hierarchy, it is well known that there is a \(\tau\)-function \(\tau^{KP}\) of the KP hierarchy such that
\[\tau^{KP}(t_1,0,t_3,0,t_5,0\ldots)=\tau^{BKP}(t_1,t_3,t_5,\ldots)^2.\]
See \cite{dateTransformationGroupsSoliton1982}. The first main result of this paper is the following theorem. For definitions and notations, see \S \ref{PRE}.
\begin{theorem}\label{CNPFBKPi}
    Let \(\tau^{BKP}({\boldsymbol{t}})\) be a \(\tau\)-function of the BKP hierarchy and \(\tau^{KP}(\boldsymbol{t})\) be the \(\tau\)-function of the KP hierarchy such that
    \[\tau^{KP}(t_1,0,t_3,0,t_5,0\ldots)=\tau^{BKP}(t_1,t_3,t_5,\ldots)^2.\]
    Define \(F^{BKP}(\boldsymbol{t}):=\log\tau^{BKP}(\boldsymbol{t})\).
    Then
    \begin{enumerate}[(1)]
        \item\label{1i} The generating series of the affine coordinates of \(\tau^{KP}(\boldsymbol{t})\) in the KP sense is given by
        \begin{equation}\label{affKPBKPi}
            \begin{aligned}
                A^{KP}(w,z)=&2\left( \sum_{m=0}^\infty (-1)^{m+1}a_{m+1,0}^{BKP}w^{-m-1}z^{-1}\right.\\+&\left.\sum_{m=0}^\infty\sum_{n=1}^\infty(-1)^{m+1}(a_{m+1,n}^{BKP}+a_{m+1,0}^{BKP}a_{0,n}^{BKP})w^{-m-1}z^{-n-1} \right)
            \end{aligned}
        \end{equation}
        where \(a_{n,m}^{BKP}\) are the affine coordinates of \(\tau^{BKP}(\boldsymbol{t})\).
        \item \label{2i} Let \(A^{BKP}\)  be the genrating series of the affine coordinates of \(\tau^{BKP}(\boldsymbol{t})\)  in the BKP sense.  The relationship between \(A^{KP}\) and \(A^{BKP}\) is given by
        \begin{equation}\label{GSKPBKPi}
            A^{BKP}(w,z)=\frac{1}{4}(zA^{KP}(w,-z)-wA^{KP}(z,-w)).
        \end{equation}
        \item \label{3i} The connected bosonic \(n\)-point functions of $F^{BKP}(\boldsymbol{t})$ are given by
        \begin{equation}\label{NPFBKPKP}
            \begin{aligned}
                &\sum_{i_1,\ldots,i_n\geq1,odd}\frac{\partial^nF^{BKP}}{\partial t_{i_1}\cdots\partial t_{i_n}}\Bigg|_{\boldsymbol{t}=0}\prod_{k=1}^nz_k^{-i_k-1}\\
                =&\frac{(-1)^{n-1}}{2^{n+1}}\sum_{\substack{\sigma:n-cycles \epsilon_1,\ldots,\\\epsilon_n\in\{\pm1\}}}\prod_{i=1}^n\hat{A}^{KP}(\epsilon_{\sigma^{i-1}(1)}z_{\sigma^{i-1}(1)},\epsilon_{\sigma^{i}(1)}z_{\sigma^{i}(1)})-\delta_{n,2}i_{z_1,z_2}\frac{z_1^2+z_2^2}{2(z_1^2-z_2^2)^2}.
            \end{aligned}
        \end{equation}
        where \(\hat{A}^{KP}\) is defined by
        \begin{equation}
            \hat{A}^{KP}(z_i,z_j)=
            \begin{cases}
                i_{z_i,z_j}\frac{1}{z_i-z_j}+A^{KP}(z_i,z_j),&i<j,\\
                A^{KP}(z_i,z_j),&i=j,\\
                i_{z_j,z_i}\frac{1}{z_i-z_j}+A^{KP}(z_i,z_j),&i>j,
            \end{cases}
        \end{equation}
        and here we use the following notation:
        \begin{align}
            i_{x,y}\frac{1}{(x+y)^m}&=\sum_{k\geq0}\binom{-m}{k}x^{-m-k}y^k,\\
            i_{x,y}\frac{x^2+y^2}{(x^2-y^2)^2}&=\sum_{n=0}^\infty(2n+1)x^{-2n-2}y^{2n}.
        \end{align}
    \end{enumerate}
\end{theorem}
Define
\begin{equation}
    \hat{A}^{BKP}(w,z)=A^{BKP}(w,z)-\frac{1}{4}-\frac{1}{2}\sum_{i=1}^\infty(-1)^iw^{-i}z^i.
\end{equation}
The formula of connected bosonic \(n\)-point functions for BKP hierarchy given by Wang and Yang \cite[Theorem 4.1]{wangBKPHierarchyAffine2022} is
\begin{equation}
    \begin{aligned}
        &\sum_{i_1,\ldots,i_n>0,odd}\frac{\partial^n \log\tau^{BKP}}{\partial t_{i_1}\cdots\partial t_{i_n}}\Bigg|_{\boldsymbol{t}=0}z_1^{-i_1}\cdots z_n^{-i_n}\\=&-\delta_{n,2}z_1z_2i_{z_1,z_2}\frac{z_1^2+z^2_2}{2(z_1^2-z_2^2)^2}+\sum_{\substack{\sigma: n-cycles,\\\epsilon_2,\ldots,\epsilon_n\in\{\pm1\}}}(-\epsilon_2\cdots\epsilon_n)\prod_{i=1}^n\xi(\epsilon_{\sigma^{i-1}(1)}z_{\sigma^{i-1}(1)},-\epsilon_{\sigma^{i}(1)}z_{\sigma^{i}(1)}),\label{NPFBKP1i}
    \end{aligned}
\end{equation}
where \(\epsilon_1=1\) and
\begin{equation}
\xi(\epsilon_{\sigma^{i-1}(1)}z_{\sigma^{i-1}(1)},-\epsilon_{\sigma^{i}(1)}z_{\sigma^{i}(1)})=
\begin{cases}
    \hat{A}^{BKP}(\epsilon_{\sigma^{i-1}(1)}z_{\sigma^{i-1}(1)},-\epsilon_{\sigma^{i}(1)}z_{\sigma^{i}(1)}),&\sigma^{i-1}(1)<\sigma^{i}(1),\\
    A^{BKP}(\epsilon_{\sigma^{i-1}(1)}z_{\sigma^{i-1}(1)},-\epsilon_{\sigma^{i}(1)}z_{\sigma^{i}(1)}),&\sigma^{i-1}(1)=\sigma^{i}(1),\\
    -\hat{A}^{BKP}(-\epsilon_{\sigma^{i}(1)}z_{\sigma^{i}(1)},\epsilon_{\sigma^{i-1}(1)}z_{\sigma^{i-1}(1)}),&\sigma^{i-1}(1)>\sigma^{i}(1).
\end{cases}
\end{equation}

The above two formulae \eqref{NPFBKP1i} and \eqref{NPFBKPKP} are different. Our second main result is that they are equivalent.
\begin{theorem}\label{EQ}
    In the above notations, we have
    \begin{equation}
        \begin{aligned}
            &\sum_{\substack{\sigma: n-cycles,\\\epsilon_2,\ldots,\epsilon_n\in\{\pm1\}}}(-\epsilon_2\cdots\epsilon_n)\prod_{i=1}^n\xi(\epsilon_{\sigma^{i-1}(1)}z_{\sigma^{i-1}(1)},-\epsilon_{\sigma^{i}(1)}z_{\sigma^{i}(1)})\\
            =&\frac{(-1)^{n-1}}{2^{n+1}}\sum_{\substack{\sigma:n-cycles\\ \epsilon_1,\ldots,\epsilon_n\in\{\pm1\}}}\prod_{i=1}^n\hat{A}^{KP}(\epsilon_{\sigma^{i-1}(1)}z_{\sigma^{i-1}(1)},\epsilon_{\sigma^{i}(1)}z_{\sigma^{i}(1)}).
        \end{aligned}
    \end{equation}
\end{theorem}

Surprisingly, the proof of this equivalence is very technical and tedious.
We will need the following technical result whose proof we put in the Appendix.
\begin{lemma}
    Let \(s(x,y)\) be a series satisfying \(s(y,x)=-s(x,y)\) and \(t(x)\) be an arbitrary series and define
\begin{align*}
    f(y,x)&=2s(y,x)+2t(y)-2t(x)+\frac{y-x}{y+x},\\
    g(y,x)&=s(y,x)+2t(y)(1-t(x))-\frac{x}{y+x}.
\end{align*}
Then we have
\begin{equation}\label{GNCi}
    \begin{aligned}
        &\sum_{\epsilon_1,\ldots,\epsilon_k\in\{\pm1\}}\sum_{\sigma\in S_k'}\prod_{i=1}^k\epsilon_{\sigma(i)}f\left( \xi_{\sigma(i)}^{\frac{1-\epsilon_{\sigma(i)}}{2}}(y_{\sigma(i)}),\xi_{\sigma(i+1)}^{\frac{1-\epsilon_{\sigma(i+1)}}{2}}(x_{\sigma(i+1)}) \right)\\
        =2^k&\sum_{\epsilon_1,\ldots,\epsilon_k\in\{\pm1\}}\sum_{\sigma\in S_k'}\prod_{i=1}^k\epsilon_{\sigma(i)}g\left( \xi_{\sigma(i)}^{\frac{1-\epsilon_{\sigma(i)}}{2}}(y_{\sigma(i)}),\xi_{\sigma(i+1)}^{\frac{1-\epsilon_{\sigma(i+1)}}{2}}(x_{\sigma(i+1)}) \right).
    \end{aligned}
\end{equation}
Here \(\xi_i\) is the nontrivial permutation of \(\{x_i,y_i\}\) and
\[S_n':=\{\sigma\in S_n\big|\sigma(1)=1\}.\]
\end{lemma}
By taking
\[s(y,x)=\sum_{m,n=1}^\infty a_{m,n}^{BKP}(y^{-m}x^{-n}-x^{-n}y^{-m}),t(x)=\sum_{m=1}^\infty a_{m,0}^{BKP}x^{-m},x_i=-y_i=z_i,\]
in \eqref{GNCi}, we prove Theorem \ref{EQ}.

The other sections of this paper are arranged as follows. In \S\ref*{PRE}, we recall some preliminaries about KP and BKP hierarchies and the formulae for the connected bosonic \(n\)-point functions of tau-functions of KP and BKP hierarchies given in \cite{zhouEmergentGeometryMirror2015} and \cite{wangBKPHierarchyAffine2022}. In \S\ref*{ComputaionBKP}, we derive our formula for the connected bosonic \(n\)-point functions of tau-functions BKP hierarchy. In \S\ref*{EFBKP}, we show the formula obtained in \S\ref*{ComputaionBKP} is equivalent to the one given by Wang and Yang in \cite{wangBKPHierarchyAffine2022}.  In \S\ref*{Conclusion}, we present some concluding remarks.

\section{Preliminaries}\label{PRE}
In this section, we give a brief review of the fermionic Fock space, the free fermions for KP hierarchy, the neutral free fermions for BKP hierarchy and their relationship. Then we recall the formulae for the connected Bosonic \(n\)-piont functions for KP and BKP hierarchies.

\subsection{The fermionic Fock space and free fermions}
First, we recall the fermionic Fock space and free fermions. See e.g. \cite[\S3]{zhouEmergentGeometryMirror2015} for more details.
\par For a sequence \(\boldsymbol{a}=(a_1,a_2,\ldots)\) of half-integers such that \(a_1<a_2<\cdots\), the elements in the set \(\{a_i\big|a_i<0\}\) will be called bubbles of \(\boldsymbol{a}\), and the elements in the set \(\left( \mathbb{Z}_{\geq0}+\frac{1}{2} \right)-\{a_1,a_2,\ldots\}\) will be called holes of \(\boldsymbol{a}\). We say \(\boldsymbol{a}\) is admissible if \(\boldsymbol{a}\) has only finitely many holes and only finitely many bubbles. For an admissible sequence \(\boldsymbol{a}\), denote
\begin{equation}
    |\boldsymbol{a}\rangle:=z^{a_1}\wedge z^{a_2}\wedge\cdots.
\end{equation}
The fermionic Fock space \(\mathcal{F}\) is the space of expressions of the form \(\sum c_{\boldsymbol{a}}|\boldsymbol{a}\rangle\) where the sum is taken over admissible sequences.
\par Define a metric on \(\mathcal{F}\) such that \(\{|\boldsymbol{a}\rangle\in\mathcal{F}\big|\boldsymbol{a}\text{ is admissible}\}\) is an orthonormal basis. We will follow notations in physics literature that the inner product of \(|v\rangle\) and \(|w\rangle\) will be denoted by \(\langle w|v\rangle\).

Let \(A\) be a Clifford algebra generated by \(\psi_i,\psi_i^*(i\in\mathbb{Z}+1/2)\) satisfying the following defining relations:
\begin{equation}
    [\psi_i,\psi_j]_+=[\psi_i^*,\psi_j^*]_+=0,[\psi_i,\psi_j^*]_+=\delta_{i+j,0}.
\end{equation}
An element of \(W=\left( \bigoplus_{i\in\mathbb{Z}}\mathbb{C}\psi_i \right)\oplus\left( \bigoplus_{i\in\mathbb{Z}}\mathbb{C}\psi_i^* \right)\) is called a free fermion. And we have a representation of the free fermions on \(\mathcal{F}\):
\begin{align}
    &\psi_r|\boldsymbol{a}\rangle=z^r\wedge|\boldsymbol{a}\rangle,\\
    &\psi_r^*|\boldsymbol{a}\rangle=\begin{cases}
        (-1)^{k-1}z^{a_1}\wedge\cdots\wedge\widehat{z^{a_k}}\wedge\cdots,&\text{if }a_k=-r\text{ for some }k,\\
        0,&\text{otherwise.}
    \end{cases}
\end{align}
The operators \(\psi_r\) and \(\psi_r^*\) are adjoint to each other.

Consider the space \(\mathcal{H}\) consisting of the Laurent series \(\sum_{n\in\mathbb{Z}}a_nz^{n-1/2}\) such that \(a_n=0\) for \(n\) large enough. Define
\[\mathcal{H}_+=\left\{\sum_{n\geq1}a_nz^{n-1/2}\in \mathcal{H}\right\},\mathcal{H}_-=\left\{\sum_{n<1}a_nz^{n-1/2}\in \mathcal{H}\right\},\]
then \(\mathcal{H}\) has a decomposition
\[\mathcal{H}=\mathcal{H}_+\oplus \mathcal{H}_-.\]
Denote the natural projection by
\[\pi_\pm:\mathcal{H}\to \mathcal{H}_\pm.\]
The big cell of Sato Grassmannian \(\operatorname{Gr}_{(0)}\) consisting of subspaces \(U\subset \mathcal{H}\) such that \(\pi_+:U\to \mathcal{H}_+\) is an isomorphism. Suppose \(U\in\operatorname{Gr}_{(0)}\) is given by a normalized basis
\[f_n=z^{n+1/2}+\sum_{m\geq0}a^{KP}_{n,m}z^{-m-1/2}, n \geq 1,\]
Zhou \cite[Theorem~3.1]{zhouEmergentGeometryMirror2015} constructed a corresponding point in the fermionic Fock space
\begin{equation}
    |U\rangle:=f_1\wedge f_2\wedge\cdots =e^{A}|0\rangle,
\end{equation}
where \(A:\mathcal{F}^{(0)}\to\mathcal{F}^{(0)}\) is given by
\begin{equation}
    A=\sum_{m,n\geq0}a^{KP}_{n,m}\psi_{-m-1/2}\psi_{-n-1/2}^*,
\end{equation}
and \(|0\rangle=z^{1/2}\wedge z^{3/2}\wedge\cdots\). 
The coefficients \(\{a^{KP}_{n,m}\}\) are called the affine coordinates  on the big cell. 
To distinguish the type B case to be discussed below, we will refer to such affine coordinates as
affine coordinates in the KP sense.
Sato \cite{satoSolitonEquationsDynamical1981} associated a tau-function of KP hierarchy to \(U\) by
\begin{equation}
    \tau^{KP}_U(t;g)=\langle0|e^{H(t)}g|0\rangle=\langle0|e^{H(t)}|U\rangle,
\end{equation}
using the boson-fermion correspondence \cite{dateTransformationGroupsSoliton1982}, where \(g=e^{A}\) and
\begin{equation}
    H(t)=\sum_{n=1}^{+\infty}t_nH_n=\sum_{n=1}^{+\infty}t_n\sum_{r\in\mathbb{Z}+\frac{1}{2}}:\psi_{-r}\psi_{r+n}^*:.
\end{equation}

\subsection{Neutral free fermions}
Here we recall neutral free fermions introduced in \cite[\S6]{jimboSolitonsInfiniteDimensional1983}.

Consider a family of operator \(\{\phi_m\}_{m\in\mathbb{Z}}\) satisfying the following anticommutative relations:
\begin{equation}
    [\phi_m,\phi_n]_+=(-1)^m\delta_{m+n,0}.
\end{equation}
For any odd integer \(n\in2\mathbb{Z}+1\), define the Hamiltonian \(H_n^B\) by
\[H_n^B=\frac{1}{2}\sum_{i\in\mathbb{Z}}(-1)^{i-1}\phi_i\phi_{-i-n},\]
and define
\[H_+(t)=\sum_{n>0:odd}t_nH_n^B.\]
These operators can be realized by actions on the fermionic Fock space $\mathcal{F}$.
If we set
\begin{equation}
    \phi_m=\frac{\psi_{-m-1/2}+(-1)^m\psi_{-m+1/2}^*}{\sqrt{2}},\hat{\phi}_m=\sqrt{-1}\frac{\psi_{-m-1/2}-(-1)^m\psi_{-m+1/2}^*}{\sqrt{2}},
\end{equation}
then
\begin{align}
    [\phi_m,\phi_n]_+&=\frac{1}{2}( [\psi_{-m-1/2},\psi_{-n-1/2}]_++(-1)^{m+n}[\psi_{-m+1/2}^*,\psi_{-n+1/2}^*]_+\\&+(-1)^m[\psi_{-m+1/2}^*,\psi_{-n-1/2}]_++(-1)^n[\psi_{-m-1/2},\psi_{-n+1/2}^*]_+ )\\
    &=(-1)^m\delta_{m+n,0}.
\end{align}
Similarly, \([\hat{\phi}_m,\hat{\phi}_n]_+=(-1)^m\delta_{m+n,0}\). And
\begin{align}
    [\phi_m,\hat{\phi}_n]_+&=\frac{\sqrt{-1}}{2}([\psi_{-m-1/2},\psi_{-n-1/2}]_+-(-1)^{m+n}[\psi_{-m+1/2}^*,\psi_{-n+1/2}^*]_+\\&-(-1)^n[\psi_{-m-1/2},\psi_{-n+1/2}^*]_++(-1)^m[\psi_{-m+1/2}^*,\psi_{-n-1/2}]_+)\\
    &=0.
\end{align}
We denote the algebra generated by \(\phi_m\)(resp. \(\hat{\phi}_m\)) by \(A'\)(resp. \(\hat{A}'\)). A tau-function of the BKP hierarchy can be represented as a Bogoliubov transform in the fermionic Fock space
\begin{equation}\label{BTBKP}
    |A\rangle:=e^{A^{BKP}}|0\rangle,
\end{equation}
where
\begin{equation}
    A^{BKP}=\sum_{n,m\geq0}a_{n,m}^{BKP}\phi_m\phi_n.
\end{equation}
Here \(a_{m,n}^{BKP}\) are called the affine coordinates in the BKP sense. 
Since \(\phi_m\phi_n=-\phi_n\phi_m\) unless \(n=m=0\) and \(\phi_0^2=\frac{1}{2}\), we can always assume
\begin{equation}
    a_{n,m}^{BKP}=-a_{m,n}^{BKP}\forall n,m\geq0.
\end{equation}
The image of \eqref{BTBKP} under the boson-fermion correspondence of type B is
\begin{equation}
    \tau^{BKP}(t;g)=\langle0|e^{H_+(t)}g|0\rangle,
\end{equation}
where \(g=e^{A^{BKP}}\). Moreover, we can define an algebra isomorphism \(\kappa:A'\to\hat{A}'\) by \(\phi_m\mapsto\hat{\phi}_m\). Then we have the following relation
\begin{equation}\label{tauKPBKP}
    \tau^{KP}(t_1,0,t_3,0,\ldots;e^{A^{BKP}+\kappa(A^{BKP})})=\tau^{BKP}(t_1,t_3,\ldots;e^{A^{BKP}})^2.
\end{equation}
See also \cite{youPolynomialSolutionsBKP1989}.

\subsection{Connected bosonic \texorpdfstring{\(n\)}{Lg}-point functions for KP hierarchy}
Let \(\tau^{KP}_U\) be the tau function of the KP hierarchy for \(|U\rangle\) and \(a^{KP}_{m,n}\) be the corresponding affine coordinates in the KP sense. 
Define a polynomial
\begin{equation}
    A^{KP}(\xi,\eta)=\sum_{m,n=0}^{+\infty}a_{m,n}^{KP}\xi^{-m-1}\eta^{-n-1}.
\end{equation}
The second author used the boson-fermion correspondence of type A to find formulae of bosoinc \(n\)-point function, and got the following theorem \cite[Theorem~5.3]{zhouEmergentGeometryMirror2015} by a combinatorial proposition \cite[Proposition~5.2]{zhouEmergentGeometryMirror2015} about determinants:
\begin{theorem}
    Let \(F^{KP}=\log\tau^{KP}_U\), then
    \begin{align}
        \sum_{i_1,\ldots,i_n\geq1}\frac{\partial^n F^{KP}}{\partial t_{i_1}\cdots\partial t_{i_n}}\Bigg|_{\boldsymbol{t}=0}z_1^{-i_1-1}\cdots z_n^{-i_n-1}&=(-1)^{n-1}\sum_{\text{n-cycles}}\prod_{i=1}^n\hat{A}^{KP}(z_{\sigma^{i-1}(1)},z_{\sigma^i(1)})-i_{z_1,z_2}\frac{\delta_{n,2}}{(z_1-z_2)^2}\label{nPFKP2}
    \end{align}
    for \(n\geq 2\) where
    \begin{equation}\label{hatA}
        \hat{A}^{KP}(z_i,z_j)=
        \begin{cases}
            i_{z_i,z_j}\frac{1}{z_i-z_j}+A^{KP}(z_i,z_j),&i<j,\\
            A^{KP}(z_i,z_j),&i=j,\\
            i_{z_j,z_i}\frac{1}{z_i-z_j}+A^{KP}(z_i,z_j),&i>j
        \end{cases}
    \end{equation}
    and
    \begin{equation}
        i_{x,y}\frac{1}{(x+y)^m}=\sum_{k\geq0}\binom{-m}{k}x^{-m-k}y^k.
    \end{equation}
\end{theorem}
\subsection{Connected bosonic \texorpdfstring{\(n\)}{Lg}-point functions for BKP hierarchy}
Let \(\tau^{BKP}_A\) be the tau function of the BKP hierarchy for \(|A\rangle\),
 and \(a^{BKP}_{n,m}\) be the corresponding affine coordinates
in the BKP sense, with  generating series  given by
\begin{align}
    A^{BKP}(w,z)=\sum_{n,m>0}(-1)^{m+n+1}a^{BKP}_{n,m}w^{-n}z^{-m}-\frac{1}{2}\sum_{n>0}(-1)^na^{BKP}_{n,0}(w^{-n}-z^{-n}).
\end{align}
Under the assumption \(a_{n,m}=-a_{m,n}\), the formula above can be rewriten as
\begin{align}\label{affBKP}
    A^{BKP}(w,z)=\frac{1}{2}\left( \sum_{n=1}^\infty\sum_{m=0}^\infty (-1)^{m+n+1}a^{BKP}_{n,m}w^{-n}z^{-m}+\sum_{n=0}^\infty\sum_{m=1}^\infty (-1)^{m+n+1} a^{BKP}_{n,m}w^{-n}z^{-m} \right)
\end{align}
and let
\begin{equation}
    \hat{A}^{BKP}(w,z)=A^{BKP}(w,z)-\frac{1}{4}-\frac{1}{2}\sum_{i=1}^\infty(-1)^iw^{-i}z^i.
\end{equation}
Wang and Yang used the boson-fermion correspondence of type B to find formulae of bosoinc \(n\)-point function, and got the following theorem \cite[Theorem~4.1]{wangBKPHierarchyAffine2022} by a combinatorial proposition \cite[Proposition~4.1]{wangBKPHierarchyAffine2022} about Pfaffians:
\begin{theorem}
    Let \(F^{BKP}(\boldsymbol{t}):=\log\tau^{BKP}_A(\boldsymbol{t})\), then
    \begin{equation}
        \begin{aligned}
            &\sum_{i_1,\ldots,i_n>0,odd}\frac{\partial^n F^{BKP}}{\partial t_{i_1}\cdots\partial t_{i_n}}\Bigg|_{\boldsymbol{t}=0}z_1^{-i_1}\cdots z_n^{-i_n}\\=&-\delta_{n,2}z_1z_2i_{z_1,z_2}\frac{z_1^2+z^2_2}{2(z_1^2-z_2^2)^2}+\sum_{\substack{\sigma: n-cycles,\\\epsilon_2,\ldots,\epsilon_n\in\{\pm1\}}}(-\epsilon_2\cdots\epsilon_n)\prod_{i=1}^n\xi(\epsilon_{\sigma^{i-1}(1)}z_{\sigma^{i-1}(1)},-\epsilon_{\sigma^{i}(1)}z_{\sigma^{i}(1)}).\label{NPFBKP1}
        \end{aligned}
    \end{equation}
where
\begin{equation}
    \xi(\epsilon_{\sigma^{i-1}(1)}z_{\sigma^{i-1}(1)},-\epsilon_{\sigma^{i}(1)}z_{\sigma^{i}(1)})=
    \begin{cases}
        \hat{A}^{BKP}(\epsilon_{\sigma^{i-1}(1)}z_{\sigma^{i-1}(1)},-\epsilon_{\sigma^{i}(1)}z_{\sigma^{i}(1)}),&\sigma^{i-1}(1)<\sigma^{i}(1),\\
        A^{BKP}(\epsilon_{\sigma^{i-1}(1)}z_{\sigma^{i-1}(1)},-\epsilon_{\sigma^{i}(1)}z_{\sigma^{i}(1)}),&\sigma^{i-1}(1)=\sigma^{i}(1),\\
        -\hat{A}^{BKP}(-\epsilon_{\sigma^{i}(1)}z_{\sigma^{i}(1)},\epsilon_{\sigma^{i-1}(1)}z_{\sigma^{i-1}(1)}),&\sigma^{i-1}(1)>\sigma^{i}(1),
    \end{cases}
\end{equation}
and we use the convention
\[i_{x,y}\frac{x^2+y^2}{(x^2-y^2)^2}=\sum_{n=0}^\infty(2n+1)x^{-2n-2}y^{2n}.\]
\end{theorem}

\section{Computation of Connected Bosonic \texorpdfstring{\(n\)}{Lg}-Point Functions for BKP Hierarchy by Using Results for KP Hierarchy}\label{ComputaionBKP}
In this section, we compute the connected bosonic \(n\)-point functions directly from \eqref{nPFKP2} and \eqref{tauKPBKP}.

Let \(F^{BKP}=\log\tau^{BKP}(t;e^{A^{BKP}})\) and \(F^{KP}=\log\tau^{KP}(t;e^{A^{BKP}+\kappa(A^{BKP})})\), then we have
\begin{equation}
    F^{BKP}(t_1,t_3,\ldots)=\frac{1}{2}F^{KP}(t_1,0,t_3,0,\ldots).
\end{equation}
Notice that
\begin{equation}
    \sum_{\epsilon_1,\ldots,\epsilon_n\in\{\pm1\}}\prod_{k=1}^n(\epsilon_kz_k)^{i_k}=\prod_{k=1}^n\sum_{\epsilon_k\in\{\pm1\}}(\epsilon_kz_k)^{i_k}=
    \begin{cases}
        2^n\prod_{k=1}^nz_k^{i_k},&\text{if }i_1,\ldots,i_n\text{ are all even},\\
        0,&\text{otherwise},
    \end{cases}
\end{equation}
for \(i_k\in\mathbb{Z}\). Define a function \(\mathcal{E}\) by
\begin{equation}
    \prod_{k=1}^nz_k^{i_k}\mapsto\sum_{\epsilon_1,\ldots,\epsilon_n\in\{\pm1\}}\prod_{k=1}^n(\epsilon_kz_k)^{i_k}.
\end{equation}
Then
\begin{equation}
    \mathcal{E}\left(\sum_{i_1,\ldots,i_n\geq1}\frac{\partial^nF^{KP}}{\partial t_{i_1}\cdots\partial t_{i_n}}\prod_{k=1}^nz_k^{-i_k-1}\right)=2^n\sum_{i_1,\ldots,i_n\geq1,odd}\frac{\partial^nF^{KP}}{\partial t_{i_1}\cdots\partial t_{i_n}}\prod_{k=1}^nz_k^{-i_k-1}.
\end{equation}
Hence
\begin{align}
    &\sum_{i_1,\ldots,i_n\geq1,odd}\frac{\partial^nF^{BKP}}{\partial t_{i_1}\cdots\partial t_{i_n}}\Bigg|_{\boldsymbol{t}=0}\prod_{k=1}^nz_k^{-i_k-1}\\
    =&\frac{1}{2^{n+1}}\mathcal{E}\left(\sum_{i_1,\ldots,i_n\geq1}\frac{\partial^nF^{KP}}{\partial t_{i_1}\cdots\partial t_{i_n}}\Bigg|_{\boldsymbol{t}=0}\prod_{k=1}^nz_k^{-i_k-1}\right)\\
    =&\frac{(-1)^{n-1}}{2^{n+1}}\sum_{\substack{\sigma:n-cycles\\ \epsilon_1,\ldots,\epsilon_n\in\{\pm1\}}}\prod_{i=1}^n\hat{A}^{KP}(\epsilon_{\sigma^{i-1}(1)}z_{\sigma^{i-1}(1)},\epsilon_{\sigma^{i}(1)}z_{\sigma^{i}(1)})-\delta_{n,2}\frac{z_1^2+z_2^2}{2(z_1^2-z_2^2)^2}.\label{NPFBKP2}
\end{align}

The formula above is in affine coordinates of a tau-function of the KP hierarchy, and we need to change them into affine coordinates of a tau-function of the BKP hierarchy. By direct computation, notice that \(a_{0,0}^{BKP}=0\), then we have
\begin{align}
    &A^{BKP}+\kappa(A^{BKP})\\
    =&\sum_{n,m\geq0}a^{BKP}_{n,m}(\phi_m\phi_n+\hat{\phi}_m\hat{\phi}_n)\\
    =&\sum_{n,m\geq0}a^{BKP}_{n,m}((-1)^n\psi_{-m-1/2}\psi_{-n+1/2}^*+(-1)^m\psi_{-m+1/2}^*\psi_{-n-1/2})\\
    =&\sum_{n,m\geq0}a^{BKP}_{n,m}((-1)^n\psi_{-m-1/2}\psi_{-n+1/2}^*+(-1)^{m+1}\psi_{-n-1/2}\psi_{-m+1/2}^*)\\
    =&2\sum_{n,m\geq0}a^{BKP}_{n,m}(-1)^n\psi_{-m-1/2}\psi_{-n+1/2}^*.
\end{align}
Here we use the relation \(a_{n,m}^{BKP}=-a_{m,n}^{BKP}\). Since \(\psi_{1/2}^*\) is not a creator, we have to do some transformation. Let
\begin{align}
    X_1&=\sum_{m=1}^\infty\sum_{n=1}^\infty2(-1)^{n}a_{n,m}^{BKP}\psi_{-m-1/2}\psi_{-n+1/2}^*,\\
    X_2&=\sum_{n=1}^\infty2(-1)^na_{n,0}^{BKP}\psi_{-1/2}\psi_{-n+1/2}^*,\\
    Y&=\sum_{m=1}^\infty2a_{0,m}^{BKP}\psi_{-m-1/2}\psi_{1/2}^*,
\end{align}
and \(X:=X_1+X_2\). Then
\begin{align}
    [X,Y]&=[X_2,Y]\\
    &=\sum_{m=1}^\infty\sum_{n=1}^\infty4(-1)^{n}a_{n,0}^{BKP}a_{0,m}^{BKP}[\psi_{-1/2}\psi_{-n+1/2}^*,\psi_{-m-1/2}\psi_{1/2}^*]\\
    &=\sum_{m=1}^\infty\sum_{n=1}^\infty4(-1)^{n+1}a_{n,0}^{BKP}a_{0,m}^{BKP}\psi_{-m-1/2}\psi_{-n+1/2}^*[\psi_{-1/2},\psi_{1/2}^*]_+\\
    &=\sum_{m=1}^\infty\sum_{n=1}^\infty4(-1)^{n+1}a_{n,0}^{BKP}a_{0,m}^{BKP}\psi_{-m-1/2}\psi_{-n+1/2}^*.
\end{align}
Notice that all the indices of the operators in \(X,[X,Y]\) are negative, we have
\[[X,[X,Y]]=0.\]
And all the indices of \(\psi\) in \([X,Y]\) are smaller than \(-1/2\), we have
\[[Y,[X,Y]]=0.\]
By the Glauber formula, we have
\[e^{X+Y}|0\rangle=e^Xe^Ye^{-[X,Y]/2}|0\rangle=e^{X-[X,Y]/2}e^Y|0\rangle=e^{X-[X,Y]/2}|0\rangle.\]
Thus,
\begin{align}
    A^{KP}&=X-\frac{[X,Y]}{2}\\
    &=2\left( \sum_{m=0}^\infty\sum_{n=1}^\infty a_{n,m}^{BKP}(-1)^{n}\psi_{-m-1/2}\psi_{-n+1/2}^*+\sum_{m=1}^\infty\sum_{n=1}^\infty(-1)^{n}a_{n,0}^{BKP}a_{0,m}^{BKP}\psi_{-m-1/2}\psi_{-n+1/2}^*\right)\\
    &=2\left( \sum_{n=0}^\infty(-1)^{n+1}a_{n+1,0}^{BKP}\psi_{-1/2}\psi^*_{-n-1/2}+\sum_{m=1}^\infty\sum_{n=0}^\infty(-1)^{n+1}(a_{n+1,m}^{BKP}+a_{n+1,0}^{BKP}a_{0,m}^{BKP})\psi_{-m-1/2}\psi^*_{-n-1/2} \right),
\end{align}
and
\begin{equation}
    \begin{aligned}
        A^{KP}(w,z)=&2\left( \sum_{m=0}^\infty (-1)^{m+1}a_{m+1,0}^{BKP}w^{-m-1}z^{-1}\right.\\+&\left.\sum_{m=0}^\infty\sum_{n=1}^\infty(-1)^{m+1}(a_{m+1,n}^{BKP}+a_{m+1,0}^{BKP}a_{0,n}^{BKP})w^{-m-1}z^{-n-1} \right).
    \end{aligned}
\end{equation}

To summarize, we have
\begin{theorem}\label{CNPFBKP}
    Let \(\tau^{BKP}({\boldsymbol{t}})\) be a \(\tau\)-function of the BKP hierarchy and \(\tau^{KP}(\boldsymbol{t})\) be the \(\tau\)-function of the KP hierarchy such that
    \[\tau^{KP}(t_1,0,t_3,0,t_5,0\ldots)=\tau^{BKP}(t_1,t_3,t_5,\ldots)^2.\]
    Define \(F^{BKP}(\boldsymbol{t}):=\log\tau^{BKP}(\boldsymbol{t})\).
    Then
    \begin{enumerate}[(1)]
        \item \label{1} The generating series of the affine coordinates of \(\tau^{KP}(\boldsymbol{t})\) in the KP sesne is given by
        \begin{equation}\label{affKPBKP}
            \begin{aligned}
                A^{KP}(w,z)=&2\left( \sum_{m=0}^\infty (-1)^{m+1}a_{m+1,0}^{BKP}w^{-m-1}z^{-1}\right.\\+&\left.\sum_{m=0}^\infty\sum_{n=1}^\infty(-1)^{m+1}(a_{m+1,n}^{BKP}+a_{m+1,0}^{BKP}a_{0,n}^{BKP})w^{-m-1}z^{-n-1} \right)
            \end{aligned}
        \end{equation}
        where \(a_{n,m}^{BKP}\) are the affine coordinates of \(\tau^{BKP}(\boldsymbol{t})\) in the BKP sense.
        \item \label{2} The relationship between \(A^{KP}\) and \(A^{BKP}\) is given by
        \begin{equation}\label{GSKPBKP}
            A^{BKP}(w,z)=\frac{1}{4}(zA^{KP}(w,-z)-wA^{KP}(z,-w)).
        \end{equation}
        \item \label{3} The connected bosonic \(n\)-point functions are given by
        \begin{equation}\label{NPFBKPKPs}
            \begin{aligned}
                &\sum_{i_1,\ldots,i_n\geq1,odd}\frac{\partial^nF^{BKP}}{\partial t_{i_1}\cdots\partial t_{i_n}}\Bigg|_{\boldsymbol{t}=0}\prod_{k=1}^nz_k^{-i_k-1}\\
                =&\frac{(-1)^{n-1}}{2^{n+1}}\sum_{\substack{\sigma:n-cycles\\ \epsilon_1,\ldots,\epsilon_n\in\{\pm1\}}}\prod_{i=1}^n\hat{A}^{KP}(\epsilon_{\sigma^{i-1}(1)}z_{\sigma^{i-1}(1)},\epsilon_{\sigma^{i}(1)}z_{\sigma^{i}(1)})-\delta_{n,2}i_{z_1,z_2}\frac{z_1^2+z_2^2}{2(z_1^2-z_2^2)^2}.
            \end{aligned}
        \end{equation}
        where \(\hat{A}^{KP}\) is defined as \eqref{hatA}.
    \end{enumerate}
\end{theorem}
\begin{proof}
    The result \ref{1} has been proved above.

    For \ref{2}, since
    \begin{equation*}
        zA^{KP}(w,-z)=2\left( \sum_{m=1}^\infty(-1)^{m+1}a_{m,0}^{BKP}w^{-m}+\sum_{m,n\geq1}(-1)^{m+n+1}(a_{m,n}^{BKP}+a_{m,0}^{BKP}a_{0,n}^{BKP})w^{-m}z^{-n} \right),
    \end{equation*}
    and
    \begin{equation*}
        wA^{KP}(z,-w)=2\left( \sum_{n=1}^\infty(-1)^{n+1}a_{n,0}^{BKP}z^{-n}+\sum_{m,n\geq1}(-1)^{m+n+1}(a_{n,m}^{BKP}+a_{n,0}^{BKP}a_{0,m}^{BKP})z^{-n}w^{-m} \right),
    \end{equation*}
    then
    \begin{align*}
        &zA^{KP}(w,-z)-wA^{KP}(z,-w)\\
        =&2\left( \sum_{m=1}^\infty\sum_{n=0}^\infty(-1)^{m+n+1}a_{m,n}w^{-m}z^{-n}+\sum_{m=0}^\infty\sum_{n=1}^\infty(-1)^{m+n+1}a_{m,n}w^{-m}z^{-n} \right)\\
        =&4A^{BKP}(w,z),
    \end{align*}
    where we use \eqref{affBKP} and \(a_{m,n}^{BKP}=-a_{n,m}^{BKP}\).

    Combine \eqref{NPFBKP2} and \eqref{affKPBKP}, we get \ref{3}.
\end{proof}
The second term on the right hand side of \eqref{NPFBKPKPs} is interesting. It also appeared in the literature about the KdV hierarchy. For example, in \cite{bertolaCorrelationFunctionsKdV2016}, the connected \(n\)-point functions for KdV hierarchy is given by
\begin{align*}
    F_{n}\left(z_{1}, \ldots, z_{n} ; \mathbf{t}\right)=-\frac{1}{n} \sum_{r \in S_{n}} \frac{\operatorname{Tr}\left(\Theta\left(z_{r_{1}}\right) \cdots \Theta\left(z_{r_{n}}\right)\right)}{\prod_{j=1}^{n}\left(z_{r_{j}}^{2}-z_{r_{j+1}}^{2}\right)}-\delta_{n, 2} \frac{z_{1}^{2}+z_{2}^{2}}{\left(z_{1}^{2}-z_{2}^{2}\right)^{2}}
\end{align*}
for some  $2\times 2$-matrix $\Theta(z_i))$.
To understand the appearance of  \(\frac{z_1^2+z_2^2}{(z_1^2-z_2^2)^2}\),
in \cite{zhouTopologicalRecursionsEynard2013}
the conventions
\begin{equation}\label{conv}
    \langle\tau_k\tau_l\rangle_0=\begin{cases}
        (-1)^k,&l=-k-1,k\geq0,\\
        (-1)^l,&k=-l-1,l\geq0,\\
        0,&\text{otherwise},
    \end{cases}
\end{equation}
has been used to define the connected \(2\)-point function in genus \(g=0\) as
\[W_{0,2}(z_1,z_2)=\frac{z_1^2+z_2^2}{(z_1^2-z_2^2)^2}.\]
 In \cite{alexandrovKdVSolvesBKP2021}, Alexandrov showed that if \(\tau=\tau(t_1,t_3,t_5,\ldots)\) is a tau-function of the KdV hierarchy, then
\[\tilde{\tau}(t_1,t_3,t_5,\ldots)=\tau\left( \frac{t_1}{2},\frac{t_3}{2},\frac{t_5}{2},\ldots \right)\]
is a tau-function of the BKP hierarchy. Thus, for odd integer \(i,j\),
\[\frac{\partial^2\log\tau}{\partial t_i\partial t_j}=4\frac{\partial^2\log\tilde{\tau}}{\partial t_i\partial t_j},\]
and by applying \eqref{NPFBKPKPs} to \(\tilde{\tau}\) we can again see the appearance of \(\frac{z_1^2+z_2^2}{(z_1^2-z_2^2)^2}\).

\section{The Equivalence of Two Fomulae of Connected \texorpdfstring{\(n\)}{Lg}-Point Functions for BKP Hierarchy}\label{EFBKP}
From the relationship \eqref{GSKPBKP}, we have
\begin{align}
    A^{BKP}(-z,z)=\frac{1}{4}\left( zA^{KP}(-z,-z)+zA^{KP}(z,z) \right),
\end{align}
i.e. we know that the formulae for 1-point functions are the same.

Before going on to the proof of the general cases
\begin{equation}\label{original}
    \begin{aligned}
        &\sum_{\substack{\sigma: n-cycles,\\\epsilon_2,\ldots,\epsilon_n\in\{\pm1\}}}(-\epsilon_2\cdots\epsilon_n)\prod_{i=1}^n\xi(\epsilon_{\sigma^{i-1}(1)}z_{\sigma^{i-1}(1)},-\epsilon_{\sigma^{i}(1)}z_{\sigma^{i}(1)})\\
        =&\frac{(-1)^{n-1}}{2^{n+1}}z_1\cdots z_n\sum_{\substack{\sigma:n-cycles\\ \epsilon_1,\ldots,\epsilon_n\in\{\pm1\}}}\prod_{i=1}^n\hat{A}^{KP}(\epsilon_{\sigma^{i-1}(1)}z_{\sigma^{i-1}(1)},\epsilon_{\sigma^{i}(1)}z_{\sigma^{i}(1)})\\
        =&-\frac{1}{2^{n+1}}\sum_{\substack{\sigma:n-cycles\\ \epsilon_1,\ldots,\epsilon_n\in\{\pm1\}}}\prod_{i=1}^n\left( \epsilon_{\sigma^i(1)}z_{\sigma^i(1)}\hat{A}^{KP}(\epsilon_{\sigma^{i-1}(1)}z_{\sigma^{i-1}(1)},\epsilon_{\sigma^{i}(1)}z_{\sigma^{i}(1)}) \right),
    \end{aligned}
\end{equation}
we introduce some notations for simplicity of presentation.

Define a map \(\mathcal{O}\)   by
\[z_1^{i_1}\cdots z_n^{i_n}\mapsto \sum_{\epsilon_1,\ldots,\epsilon_n\in\{\pm1\}}\prod_{k=1}^{n}\epsilon_k(\epsilon_kz_k)^{i_k}=\begin{cases}
    2^n\prod_{k=1}^nz_k^{i_k},&i_1,\ldots,i_n\text{ are all odd},\\
    0,&\text{otherwise},
\end{cases}\]
and a closely related map  \(\tilde{\mathcal{O}}\) by
\[z_1^{i_1}\cdots z_n^{i_n}\mapsto z_1^{i_1}\sum_{\epsilon_2,\ldots,\epsilon_n\in\{\pm1\}}\prod_{k=2}^{n}\epsilon_k(\epsilon_kz_k)^{i_k}=\begin{cases}
    2^{n-1}\prod_{k=1}^nz_k^{i_k},&i_2,\ldots,i_n\text{ are all odd},\\
    0,&\text{otherwise}.
\end{cases}\]
And from now on, for convinience, we will omit \(i_{x,y}\) in the previous convention, i.e. all the rational functions of the form \(\frac{a_j}{b_i+a_j}\) represent a series as  follows:
\begin{align*}
    \frac{a_j}{b_i+a_j}=\begin{cases}
        -\sum_{n=0}^\infty(-b_i)^{-n-1}a_j^{n+1},&i<j,\\
        0,&i=j,\\
        \sum_{n=0}^\infty(-a_j)^{-n}b_i^n,&i>j,
    \end{cases}
\end{align*}
and let
\[\frac{b_i-a_j}{b_i+a_j}=\frac{b_i}{a_j+b_i}-\frac{a_j}{b_i+a_j}.\]
For example,
\begin{align*}
    \frac{z_j}{z_i-z_j}&=\begin{cases}
        \sum_{n=0}^\infty z_i^{-n-1}z_j^{n+1},&i<j,\\
        0,&i=j,\\
        -\sum_{n=0}^\infty z_j^{-n}z_i^n,&i>j,
    \end{cases}\\
    \frac{z_i+z_j}{z_i-z_j}&=\begin{cases}
        1+2\sum_{n=0}^\infty z_i^{-n-1}z_j^{n+1},&i<j,\\
        0,&i=j,\\
        -1-2\sum_{n=0}^\infty z_j^{-n-1}z_i^{n+1},&i>j,
    \end{cases}\\
    \frac{y_i-x_j}{y_i+x_j}&=\begin{cases}
        1+2\sum_{n=0}^\infty (-y_i)^{-n-1}x_j^{n+1},&i<j,\\
        0,&i=j,\\
        -1-2\sum_{n=0}^\infty (-x_j)^{-n-1}y_i^{n+1},&i>j,
    \end{cases}
\end{align*}
With these notations, \eqref{original} becomes
\begin{equation} \label{eqn:tO-O}
\begin{split}
    &\tilde{\mathcal{O}}\left( \sum_{\sigma\in S_n'}\prod_{i=1}^n\left( \sum_{m=1}^\infty\sum_{n'=0}^\infty 2a_{m,n'}^{BKP}((-z_{\sigma(i)})^{-m}z_{\sigma(i+1)}^{-n'}-(-z_{\sigma(i)})^{-n'}z_{\sigma(i+1)}^{-m})+\frac{z_{\sigma(i)}+z_{\sigma(i+1)}}{z_{\sigma(i)}-z_{\sigma(i+1)}} \right) \right) \\
    =&2^{n-1}\cdot \mathcal{O}\left( \sum_{\sigma\in S'_n}\prod_{i=1}^n \left( \sum_{m=1}^\infty\sum_{n'=0}^\infty2(a_{m,n'}^{BKP}+a_{m,0}^{BKP}a_{0,n'}^{BKP})(-z_{\sigma(i)})^{-m}z_{\sigma(i+1)}^{-n'}+\frac{z_{\sigma(i+1)}}{z_{\sigma(i)}-z_{\sigma(i+1)}} \right)\right)
\end{split}
\end{equation}
Here we use the convention \(\sigma(n+1)=\sigma(1)\) and $S_n'$ is the group of permutations $\sigma$ in $S_n$ such that $\sigma(1) =1$.
Note the two-sides of \eqref{eqn:tO-O} use two different averaging operations $\tilde{\mathcal O}$ and ${\mathcal O}$, to show that they equal to each other,
the first step is to rewrite the left-hand side also by ${\mathcal O}$.
Denote the left hand side of \eqref{eqn:tO-O} by \(L\), then
\begin{equation} \label{eqn:L=-L}
L(z_1,z_2,\ldots,z_n)=-L(-z_1,z_2,\ldots,z_n).
\end{equation}
To see this, for \(\sigma\in S_n'\), we denote
\[l_{\sigma}(z_1,\ldots,z_n):=\prod_{i=1}^n\left( \sum_{m=1}^\infty\sum_{n'=0}^\infty 2a_{m,n'}^{BKP}((-z_{\sigma(i)})^{-m}z_{\sigma(i+1)}^{-n'}-(-z_{\sigma(i)})^{-n'}z_{\sigma(i+1)}^{-m})+\frac{z_{\sigma(i)}+z_{\sigma(i+1)}}{z_{\sigma(i)}-z_{\sigma(i+1)}} \right),\]
and we have the following symmetry property:
\[l_{\sigma}(z_1,\ldots,z_n)=-(-1)^{n-1}l_{\tilde{\sigma}}(-z_1,\ldots,-z_n),\]
where \(\tilde{\sigma}\in S_n'\) is defined by
\[\tilde{\sigma}(i)=\sigma(n+2-i),i=2,\ldots,n.\]
Note \(\tilde{O}\) is linear, we have
\begin{align*}
    L(z_1,\ldots,z_n)&= \sum_{\sigma\in S_n'} \tilde{\mathcal{O}}(l_{\sigma}(z_1,\ldots,z_n))\\
    &=-\sum_{\tilde{\sigma}\in S_n'} \tilde{\mathcal{O}}((-1)^{n-1}l_{\tilde{\sigma}}(-z_1,\ldots,-z_n))\\
    &=-\sum_{\tilde{\sigma}\in S_n'} \tilde{\mathcal{O}}(l_{\tilde{\sigma}}(-z_1,z_2,\ldots,z_n))\\
    &=-L(-z_1,z_2,\ldots,z_n)
\end{align*}
Here we use the property that for any function \(f\),
\[\tilde{\mathcal{O}}(f(z_1,\ldots,z_n))=\tilde{\mathcal{O}}((-1)^{n-1}f(z_1,-z_2,\ldots,-z_n)).\]

Now we use \eqref{eqn:L=-L} to to see that \eqref{eqn:tO-O} is equivalent to
\begin{equation}\label{T1}
    \begin{aligned}
        &\mathcal{O}\left( \sum_{\sigma\in S_n'}\prod_{i=1}^n\left( \sum_{m=1}^\infty\sum_{n'=0}^\infty 2a_{m,n'}^{BKP}((-z_{\sigma(i)})^{-m}z_{\sigma(i+1)}^{-n'}-(-z_{\sigma(i)})^{-n'}z_{\sigma(i+1)}^{-m})+\frac{z_{\sigma(i)}+z_{\sigma(i+1)}}{z_{\sigma(i)}-z_{\sigma(i+1)}} \right) \right)\\
        =&L(z_1,\ldots,z_n)-L(-z_1,z_2,\ldots,z_n)=2L(z_1,\ldots,z_n)\\
        =&2^{n}\cdot \mathcal{O}\left( \sum_{\sigma\in S'_n}\prod_{i=1}^n \left( \sum_{m=1}^\infty\sum_{n'=0}^\infty2(a_{m,n'}^{BKP}+a_{m,0}^{BKP}a_{0,n'}^{BKP})(-z_{\sigma(i)})^{-m}z_{\sigma(i+1)}^{-n'}+\frac{z_{\sigma(i+1)}}{z_{\sigma(i)}-z_{\sigma(i+1)}} \right)\right).
    \end{aligned}
\end{equation}
Denote
\[ s_0(w,z):=\sum_{m=1}^\infty\sum_{n'=1}^\infty a_{m,n'}^{BKP}((-w)^{-m}z^{-n'}-(-w)^{-n'}z^{-m}), \]
and
\[t_0(z)=\sum_{m=1}^\infty a_{m,0}^{BKP}z^{-m},\]
\eqref{T1} becomes
\begin{equation}\label{st}
    \begin{aligned}
        &\mathcal{O}\left( \sum_{\sigma\in S_n'}\prod_{i=1}^n\left( 2(s_0(z_{\sigma(i)},z_{\sigma(i+1)})+t_0(-z_{\sigma(i)})-t_0(z_{\sigma(i+1)}))+\frac{z_{\sigma(i)}+z_{\sigma(i+1)}}{z_{\sigma(i)}-z_{\sigma(i+1)}} \right) \right)\\
        =&2^n\mathcal{O}\left( \sum_{\sigma\in S_n'}\prod_{i=1}^n\left( s_0(z_{\sigma(i)},z_{\sigma(i+1)})+2t_0(-z_{\sigma(i)})(1-t_0(z_{\sigma(i+1)}))+\frac{z_{\sigma(i+1)}}{z_{\sigma(i)}-z_{\sigma(i+1)}} \right) \right).
    \end{aligned}
\end{equation}
We will need the following technical lemma.
\begin{lemma}\label{GEQ}
    Let \(s(x,y)\) be a series satisfying \(s(y,x)=-s(x,y)\) and \(t(x)\) be an arbitrary series and define
\begin{align*}
    f(y,x)&=2s(y,x)+2t(y)-2t(x)+\frac{y-x}{y+x},\\
    g(y,x)&=s(y,x)+2t(y)(1-t(x))-\frac{x}{y+x}.
\end{align*}
Then we have
\begin{subequations}\label{GNC}
    \begin{align}
        &\sum_{\epsilon_1,\ldots,\epsilon_k\in\{\pm1\}}\sum_{\sigma\in S_k'}\prod_{i=1}^k\epsilon_{\sigma(i)}f\left( \xi_{\sigma(i)}^{\frac{1-\epsilon_{\sigma(i)}}{2}}(y_{\sigma(i)}),\xi_{\sigma(i+1)}^{\frac{1-\epsilon_{\sigma(i+1)}}{2}}(x_{\sigma(i+1)}) \right)\label{LHS}\\
        =2^k&\sum_{\epsilon_1,\ldots,\epsilon_k\in\{\pm1\}}\sum_{\sigma\in S_k'}\prod_{i=1}^k\epsilon_{\sigma(i)}g\left( \xi_{\sigma(i)}^{\frac{1-\epsilon_{\sigma(i)}}{2}}(y_{\sigma(i)}),\xi_{\sigma(i+1)}^{\frac{1-\epsilon_{\sigma(i+1)}}{2}}(x_{\sigma(i+1)})\label{RHS} \right).
    \end{align}
\end{subequations}
Here \(\xi_i\) is the nontrivial permutation of \(\{x_i,y_i\}\).
\end{lemma}
For the proof of this lemma, see appendix \ref{Proof}. Set \(s(w,z)=s_0(-w,z)\) and \(t(z)=t_0(z)\) and \(x_i=-y_i=z_i\), we can obtain \eqref{st}. To summarize,
we have completed the proof of Theorem \ref{EQ} which we restate here:

\begin{theorem}
   In the above notations, we have
    \begin{equation}
        \begin{aligned}
            &\sum_{\substack{\sigma: n-cycles,\\\epsilon_2,\ldots,\epsilon_n\in\{\pm1\}}}(-\epsilon_2\cdots\epsilon_n)\prod_{i=1}^n\xi(\epsilon_{\sigma^{i-1}(1)}z_{\sigma^{i-1}(1)},-\epsilon_{\sigma^{i}(1)}z_{\sigma^{i}(1)})\\
            =&\frac{(-1)^{n-1}}{2^{n+1}}\sum_{\substack{\sigma:n-cycles\\ \epsilon_1,\ldots,\epsilon_n\in\{\pm1\}}}\prod_{i=1}^n\hat{A}^{KP}(\epsilon_{\sigma^{i-1}(1)}z_{\sigma^{i-1}(1)},\epsilon_{\sigma^{i}(1)}z_{\sigma^{i}(1)}).
        \end{aligned}
    \end{equation}
\end{theorem}

\section{Conclusion}\label{Conclusion}

In \cite{jimboSolitonsInfiniteDimensional1983} we have seen that
many types of integrable hierarchies are reductions of KP hierarchy, such as KdV, BKP, CKP, DKP hierarchies and so on.
It was proposed in \cite{zhouFermionicComputationsIntegrable2015} to compute the $n$-point functions
by fermionic approach
for such reductions from the $n$-point functions of KP hierarchy in \cite{zhouEmergentGeometryMirror2015}.
In this paper we have carried out this proposal for the BKP hierarchy,
and have shown the result matches with the formula by Wang and Yang \cite{wangBKPHierarchyAffine2022} by different method.

Our approach clearly is applicable to other reductions of the KP hierarchy.
For example,
Wang and Yang have derived formulae for
the $n$-point function for 2-Toda hierarchy \cite{wang2023diagonal} and 2-BKP hierarchy \cite{wang2023connected} based on the corresponding boson-fermion correspondence.
It is natural to check that they match with the formulae derived using our approach.
As our result in this paper indicates,
such checking can be quite involved.
It is also interesting to check  that the formulae using our approach
match with the formulae obtained by Bertola-Dubrovin-Yang approach \cite{bertolaCorrelationFunctionsKdV2016,bertola2021simple}.
Again this can be technically complicated. We leave the investigations of such problems to future research.

\appendix

\section{Proof of Lemma \ref{GEQ}}\label{Proof}
Now we prove Lemma \ref{GEQ} by brute force.
At first, it's trivial that
\begin{align*}
    &\left( 2s(y_i,x_j)+2t(y_i)-2t(x_j)+\frac{y_i-x_j}{y_i+x_j} \right)-\left( 2s(x_j,y_i)+2t(x_j)-2t(y_i)+\frac{x_j-y_i}{x_j+y_i} \right)\\
    =&2\left( \left( s(y_i,x_j)+2t(y_i)(1-t(x_j))-\frac{x_j}{y_i+x_j} \right)-\left( s(x_j,y_i)+2t(x_j)(1-2t(y_i))-\frac{y_i}{x_j+y_i} \right) \right)
\end{align*}
for arbitrary \(i,j\). Particularly, set \(i=j=1\) and we know \eqref{GNC} holds for \(k=1\).

We will use induction to prove this equality. Assume that this equality holds for \(n<k\), and denote \eqref{LHS} and \eqref{RHS} by \(W(x_1,y_1,\ldots,x_k,y_k)\) and \(2^kZ(x_1,y_1,\ldots,x_k,y_k)\) respectively. Notice that
\[f(y,x)=-f(x,y),\]
then
\begin{align*}
    &W(x_1,y_1,\ldots,x_k,y_k)\\
    =&\sum_{\epsilon_1,\ldots,\epsilon_k\in\{\pm1\}}\sum_{\sigma\in S_k'}\prod_{i=1}^k\epsilon_{\sigma(i)}f\left( \xi_{\sigma(i)}^{\frac{1-\epsilon_{\sigma(i)}}{2}}(y_{\sigma(i)}),\xi_{\sigma(i+1)}^{\frac{1-\epsilon_{\sigma(i+1)}}{2}}(x_{\sigma(i+1)}) \right)\\
    =&\sum_{\epsilon_1=1,\epsilon_2,\ldots,\epsilon_k\in\{\pm1\}}\sum_{\sigma\in S_k'}\prod_{i=1}^{k-1}\epsilon_{\sigma(i+1)}f\left( \xi_{\sigma(i)}^{\frac{1-\epsilon_{\sigma(i)}}{2}}(y_{\sigma(i)}),\xi_{\sigma(i+1)}^{\frac{1-\epsilon_{\sigma(i+1)}}{2}}(x_{\sigma(i+1)}) \right)\cdot f\left( \xi_{\sigma(k)}^{\frac{1-\epsilon_{\sigma(k)}}{2}}(y_{\sigma(k)}),x_1 \right)\\
    -&\sum_{\epsilon_1=-1,\epsilon_2,\ldots,\epsilon_k\in\{\pm1\}}\sum_{\sigma\in S_k'}f\left( x_1,\xi_{\sigma(2)}^{\frac{1-\epsilon_{\sigma(2)}}{2}}(x_{\sigma(2)}) \right)\cdot \prod_{i=2}^{k}\epsilon_{\sigma(i)}f\left( \xi_{\sigma(i)}^{\frac{1-\epsilon_{\sigma(i)}}{2}}(y_{\sigma(i)}),\xi_{\sigma(i+1)}^{\frac{1-\epsilon_{\sigma(i+1)}}{2}}(x_{\sigma(i+1)}) \right)\\
    =&\sum_{l=2}^k\left( \sum_{\substack{\epsilon_1=1,\epsilon_l=1,\\\epsilon_2,\ldots,\hat{\epsilon}_l,\ldots,\epsilon_k\in\{\pm1\}}}\sum_{\substack{\sigma\in S_k\\ \sigma(1)=1,\sigma(k)=l}}\prod_{i=1}^{k-1}\epsilon_{\sigma(i+1)}f\left( \xi_{\sigma(i)}^{\frac{1-\epsilon_{\sigma(i)}}{2}}(y_{\sigma(i)}),\xi_{\sigma(i+1)}^{\frac{1-\epsilon_{\sigma(i+1)}}{2}}(x_{\sigma(i+1)}) \right)\cdot f\left( y_l,x_1 \right) \right.\\
    &+\sum_{\substack{\epsilon_1=1,\epsilon_l=-1,\\\epsilon_2,\ldots,\hat{\epsilon}_l,\ldots,\epsilon_k\in\{\pm1\}}}\sum_{\substack{\sigma\in S_k\\ \sigma(1)=1,\sigma(k)=l}}\prod_{i=1}^{k-1}\epsilon_{\sigma(i+1)}f\left( \xi_{\sigma(i)}^{\frac{1-\epsilon_{\sigma(i)}}{2}}(y_{\sigma(i)}),\xi_{\sigma(i+1)}^{\frac{1-\epsilon_{\sigma(i+1)}}{2}}(x_{\sigma(i+1)}) \right)\cdot f\left( x_l,x_1 \right)\\
    &-\sum_{\substack{\epsilon_1=-1,\epsilon_l=1,\\\epsilon_2,\ldots,\hat{\epsilon}_l,\ldots,\epsilon_k\in\{\pm1\}}}\sum_{\substack{\sigma\in S_k,\\\sigma(1)=1,\sigma(2)=l}}f\left( x_1,x_l \right)\cdot \prod_{i=2}^{k}\epsilon_{\sigma(i)}f\left( \xi_{\sigma(i)}^{\frac{1-\epsilon_{\sigma(i)}}{2}}(y_{\sigma(i)}),\xi_{\sigma(i+1)}^{\frac{1-\epsilon_{\sigma(i+1)}}{2}}(x_{\sigma(i+1)}) \right)\\
    &-\left. \sum_{\substack{\epsilon_1=-1,\epsilon_l=-1,\\\epsilon_2,\ldots,\hat{\epsilon}_l,\ldots,\epsilon_k\in\{\pm1\}}}\sum_{\substack{\sigma\in S_k,\\\sigma(1)=1,\sigma(2)=l}}f\left( x_1,y_l \right)\cdot \prod_{i=2}^{k}\epsilon_{\sigma(i)}f\left( \xi_{\sigma(i)}^{\frac{1-\epsilon_{\sigma(i)}}{2}}(y_{\sigma(i)}),\xi_{\sigma(i+1)}^{\frac{1-\epsilon_{\sigma(i+1)}}{2}}(x_{\sigma(i+1)}) \right) \right).
\end{align*}
Denote \(\tilde{\epsilon}_i=-\epsilon_{k+2-i}\) and for any \(\sigma\in S_k'\) we define \(\tilde{\sigma}\in S_k'\) as a permutaion such that
\[\tilde{\sigma}(i)=\sigma(k+2-i) \quad \forall i=2,\ldots,k.\]
Since
\begin{align*}
    f\left( \xi_{\sigma(i)}^{\frac{1-\epsilon_{\sigma(i)}}{2}}(y_{\sigma(i)}),\xi_{\sigma(i+1)}^{\frac{1-\epsilon_{\sigma(i+1)}}{2}}(x_{\sigma(i+1)}) \right)&=f\left( \xi_{\sigma(i)}^{\frac{\tilde{\epsilon}_{\sigma(i)}-1}{2}}(x_{\sigma(i)}),\xi_{\sigma(i+1)}^{\frac{\tilde{\epsilon}_{\sigma(i+1)}-1}{2}}(y_{\sigma(i+1)}) \right)\\
    &=-f\left( \xi_{\tilde{\sigma}(k+1-i)}^{\frac{1-\tilde{\epsilon}_{\tilde{\sigma}(k+1-i)}}{2}}(y_{\tilde{\sigma}(k+1-i)}),\xi_{\tilde{\sigma}(k+2-i)}^{\frac{1-\tilde{\epsilon}_{\tilde{\sigma}(k+2-i)}}{2}}(y_{\tilde{\sigma}(k+2-i)}) \right),
\end{align*}
we have
\begin{align*}
    &\sum_{\substack{\epsilon_1=1,\epsilon_l=1,\\\epsilon_2,\ldots,\hat{\epsilon}_l,\ldots,\epsilon_k\in\{\pm1\}}}\sum_{\substack{\sigma\in S_k\\ \sigma(1)=1,\sigma(k)=l}}\prod_{i=1}^{k-1}\epsilon_{\sigma(i+1)}f\left( \xi_{\sigma(i)}^{\frac{1-\epsilon_{\sigma(i)}}{2}}(y_{\sigma(i)}),\xi_{\sigma(i+1)}^{\frac{1-\epsilon_{\sigma(i+1)}}{2}}(x_{\sigma(i+1)}) \right)\cdot f\left( y_l,x_1 \right)\\
    -&\sum_{\substack{\epsilon_1=-1,\epsilon_l=-1,\\\epsilon_2,\ldots,\hat{\epsilon}_l,\ldots,\epsilon_k\in\{\pm1\}}}\sum_{\substack{\sigma\in S_k,\\\sigma(1)=1,\sigma(2)=l}}f\left( x_1,y_l \right)\cdot \prod_{i=2}^{k}\epsilon_{\sigma(i)}f\left( \xi_{\sigma(i)}^{\frac{1-\epsilon_{\sigma(i)}}{2}}(y_{\sigma(i)}),\xi_{\sigma(i+1)}^{\frac{1-\epsilon_{\sigma(i+1)}}{2}}(x_{\sigma(i+1)}) \right)\\
    =&\sum_{\substack{\epsilon_1=1,\epsilon_l=1,\\\epsilon_2,\ldots,\hat{\epsilon}_l,\ldots,\epsilon_k\in\{\pm1\}}}\sum_{\substack{\sigma\in S_k\\ \sigma(1)=1,\sigma(k)=l}}\prod_{i=1}^{k-1}\epsilon_{\sigma(i+1)}f\left( \xi_{\sigma(i)}^{\frac{1-\epsilon_{\sigma(i)}}{2}}(y_{\sigma(i)}),\xi_{\sigma(i+1)}^{\frac{1-\epsilon_{\sigma(i+1)}}{2}}(x_{\sigma(i+1)}) \right)\cdot f\left( y_l,x_1 \right)\\
    -&\sum_{\substack{\epsilon_1=-1,\epsilon_l=-1,\\\epsilon_2,\ldots,\hat{\epsilon}_l,\ldots,\epsilon_k\in\{\pm1\}}}\sum_{\substack{\sigma\in S_k\\ \sigma(1)=1,\sigma(k)=l}}\prod_{i=1}^{k-1}\epsilon_{\sigma(i+1)}f\left( \xi_{\sigma(i)}^{\frac{1-\epsilon_{\sigma(i)}}{2}}(y_{\sigma(i)}),\xi_{\sigma(i+1)}^{\frac{1-\epsilon_{\sigma(i+1)}}{2}}(x_{\sigma(i+1)}) \right)\cdot f\left( y_l,x_1 \right)\\
    =&W(x_l,y_1,x_2,y_2,\ldots,\hat{x}_l,\hat{y}_l,\ldots,x_k,y_k)f(y_l,x_1).
\end{align*}
Similarly, we have
\begin{align*}
    &\sum_{\substack{\epsilon_1=1,\epsilon_l=-1,\\\epsilon_2,\ldots,\hat{\epsilon}_l,\ldots,\epsilon_k\in\{\pm1\}}}\sum_{\substack{\sigma\in S_k\\ \sigma(1)=1,\sigma(k)=l}}\prod_{i=1}^{k-1}\epsilon_{\sigma(i+1)}f\left( \xi_{\sigma(i)}^{\frac{1-\epsilon_{\sigma(i)}}{2}}(y_{\sigma(i)}),\xi_{\sigma(i+1)}^{\frac{1-\epsilon_{\sigma(i+1)}}{2}}(x_{\sigma(i+1)}) \right)\cdot f\left( x_l,x_1 \right)\\
    -&\sum_{\substack{\epsilon_1=-1,\epsilon_l=1,\\\epsilon_2,\ldots,\hat{\epsilon}_l,\ldots,\epsilon_k\in\{\pm1\}}}\sum_{\substack{\sigma\in S_k,\\\sigma(1)=1,\sigma(2)=l}}f\left( x_1,x_l \right)\cdot \prod_{i=2}^{k}\epsilon_{\sigma(i)}f\left( \xi_{\sigma(i)}^{\frac{1-\epsilon_{\sigma(i)}}{2}}(y_{\sigma(i)}),\xi_{\sigma(i+1)}^{\frac{1-\epsilon_{\sigma(i+1)}}{2}}(x_{\sigma(i+1)}) \right)\\
    =&-W(y_l,y_1,x_2,y_2,\ldots,\hat{x}_l,\hat{y}_l,\ldots,x_k,y_k)f(x_l,x_1).
\end{align*}
Hence
\begin{subequations}\label{IW}
    \begin{align}
        &W(x_1,y_1,\ldots,x_k,y_k)\\
        =&\sum_{l=2}^k\left( W(x_l,y_1,x_2,y_2,\ldots,\hat{x}_l,\hat{y}_l,\ldots,x_k,y_k)f(y_l,x_1)\right.\notag\\
        &\left.-W(y_l,y_1,x_2,y_2,\ldots,\hat{x}_l,\hat{y}_l,\ldots,x_k,y_k)f(x_l,x_1) \right)
    \end{align}
\end{subequations}
Denote
\begin{equation}\label{tZ}
    \tilde{Z}(x_1,y_1,\ldots,x_k,y_k)=\sum_{\epsilon_1=1,\epsilon_2,\ldots,\epsilon_k\in\{\pm1\}}\sum_{\sigma\in S_k'}\prod_{i=1}^k\epsilon_{\sigma(i)}g\left( \xi_{\sigma(i)}^{\frac{1-\epsilon_{\sigma(i)}}{2}}(y_{\sigma(i)}),\xi_{\sigma(i+1)}^{\frac{1-\epsilon_{\sigma(i+1)}}{2}}(x_{\sigma(i+1)}) \right),
\end{equation}
and calculate \(Z\) in the same way, we can obtain
\begin{subequations}
    \begin{align}
        &Z(x_1,y_1,\ldots,x_k,y_k)\\
        =&\sum_{l=2}^k\left( \tilde{Z}(x_l,y_1,x_2,y_2,\ldots,\hat{x}_l,\hat{y}_l,\ldots,x_k,y_k)g(y_l,x_1) \right.\notag\\
        &+\tilde{Z}(y_1,x_l,x_2,y_2,\ldots,\hat{x}_l,\hat{y}_l,\ldots,x_k,y_k)g(x_1,y_l)\notag\\
        &-\tilde{Z}(y_l,y_1,x_2,y_2,\ldots,\hat{x}_l,\hat{y}_l,\ldots,x_k,y_k)g(x_l,x_1)\notag\\
        &\left.-\tilde{Z}(y_1,y_l,x_2,y_2,\ldots,\hat{x}_l,\hat{y}_l,\ldots,x_k,y_k)g(x_1,x_l)\right).
    \end{align}
\end{subequations}
According to the inductive assumptions and
\[Z(x_1,y_1,\ldots,x_k,y_k)=\tilde{Z}(x_1,y_1,x_2,y_2\ldots,x_k,y_k)-\tilde{Z}(y_1,x_1,x_2,y_2\ldots,x_k,y_k),\]
and
\[f(y,x)=g(y,x)-g(x,y),\]
we have
\begin{align*}
    &W(x_1,y_1,\ldots,x_k,y_k)-2^kZ(x_1,y_1,\ldots,x_k,y_k)\\
    =&\sum_{l=2}^k\left(2^{k-1}\left( Z(x_l,y_1,x_2,y_2,\ldots,\hat{x}_l,\hat{y}_l,\ldots,x_k,y_k)f(y_l,x_1)\right.\right.\\&\left.-Z(y_l,y_1,x_2,y_2,\ldots,\hat{x}_l,\hat{y}_l,\ldots,x_k,y_k)f(x_l,x_1) \right)\\
    &-2^k\left( \tilde{Z}(x_l,y_1,x_2,y_2,\ldots,\hat{x}_l,\hat{y}_l,\ldots,x_k,y_k)g(y_l,x_1) \right.\\
    &+\tilde{Z}(y_1,x_l,x_2,y_2,\ldots,\hat{x}_l,\hat{y}_l,\ldots,x_k,y_k)g(x_1,y_l)\\
    &-\tilde{Z}(y_l,y_1,x_2,y_2,\ldots,\hat{x}_l,\hat{y}_l,\ldots,x_k,y_k)g(x_l,x_1)\\
    &\left.-\tilde{Z}(y_1,y_l,x_2,y_2,\ldots,\hat{x}_l,\hat{y}_l,\ldots,x_k,y_k)g(x_1,x_l)\right)\\
    =&\sum_{l=2}^k\left( \tilde{Z}(y_l,y_1,x_2,y_2,\ldots,\hat{x}_l,\hat{y}_l,\ldots,x_k,y_k)+\tilde{Z}(y_1,y_l,x_2,y_2,\ldots,\hat{x}_l,\hat{y}_l,\ldots,x_k,y_k) \right)\cdot\\
    &\cdot\left( g(x_1,x_l)+g(x_l,x_1) \right)\\
    &-\left( \tilde{Z}(x_l,y_1,x_2,y_2,\ldots,\hat{x}_l,\hat{y}_l,\ldots,x_k,y_k)+\tilde{Z}(y_1,x_l,x_2,y_2,\ldots,\hat{x}_l,\hat{y}_l,\ldots,x_k,y_k) \right)\cdot\\
    &\cdot\left( g(x_1,y_l)+g(y_l,x_1) \right)\\
    =&\sum_{l=2}^k\left( \tilde{Z}(x_l,y_1,x_2,y_2,\ldots,\hat{x}_l,\hat{y}_l,\ldots,x_k,y_k)+\tilde{Z}(y_1,x_l,x_2,y_2,\ldots,\hat{x}_l,\hat{y}_l,\ldots,x_k,y_k) \right)\cdot\\
    &\cdot(2t(y_l)-1)(2t(x_1)-1)\\
    &-\left( \tilde{Z}(y_l,y_1,x_2,y_2,\ldots,\hat{x}_l,\hat{y}_l,\ldots,x_k,y_k)+\tilde{Z}(y_1,y_l,x_2,y_2,\ldots,\hat{x}_l,\hat{y}_l,\ldots,x_k,y_k) \right)\cdot\\
    &\cdot(2t(x_l)-1)(2t(x_1)-1).
\end{align*}
Denote that
\begin{align*}
    &F(y_1,x_2,y_2,\ldots,x_k,y_k)\\
    =\sum_{l=2}^k&\left( \tilde{Z}(x_l,y_1,x_2,y_2,\ldots,\hat{x}_l,\hat{y}_l,\ldots,x_k,y_k)+\tilde{Z}(y_1,x_l,x_2,y_2,\ldots,\hat{x}_l,\hat{y}_l,\ldots,x_k,y_k) \right)(2t(y_l)-1)\\
    -&\left( \tilde{Z}(y_l,y_1,x_2,y_2,\ldots,\hat{x}_l,\hat{y}_l,\ldots,x_k,y_k)+\tilde{Z}(y_1,y_l,x_2,y_2,\ldots,\hat{x}_l,\hat{y}_l,\ldots,x_k,y_k) \right)(2t(x_l)-1).
\end{align*}
It's enough to prove that
\[F(y_1,x_2,y_2,\ldots,x_k,y_k)=0.\]
Note for \(n<k\), by induction, we have
\begin{equation}\label{Fn0}
    F(y_1,x_2,y_2,\ldots,x_n,y_n)=0.
\end{equation}

In order to "split" the sum
\[\tilde{Z}(x_l,y_1,x_2,y_2,\ldots,\hat{x}_l,\hat{y}_l,\ldots,x_k,y_k)+\tilde{Z}(y_1,x_l,x_2,y_2,\ldots,\hat{x}_l,\hat{y}_l,\ldots,x_k,y_k),\]
we compute
\begin{subequations}\label{Rsplit}
    \begin{align}
        &\tilde{Z}(x_1,y_1,x_2,y_2,\ldots,x_k,y_k)+\tilde{Z}(y_1,x_1,x_2,y_2,\ldots,x_k,y_k)\\
        =&\tilde{Z}(x_1,y_1,x_2,y_2,\ldots,x_k,y_k)+(-1)^{k-1}\tilde{Z}(y_1,x_1,y_k,x_k,\ldots,y_2,x_2)\\
        =&\sum_{\epsilon_1=1,\epsilon_2,\ldots,\epsilon_k\in\{\pm1\}}\sum_{\sigma\in S_k'}\left( \prod_{i=1}^k\epsilon_{\sigma(i)} \right)\left( \prod_{i=1}^kg\left( \xi_{\sigma(i)}^{\frac{1-\epsilon_{\sigma(i)}}{2}}(y_{\sigma(i)}),\xi_{\sigma(i+1)}^{\frac{1-\epsilon_{\sigma(i+1)}}{2}}(x_{\sigma(i+1)}) \right) \right.\notag\\
        &\left.+(-1)^{k-1}\prod_{i=1}^kg\left( \xi_{\sigma(i+1)}^{\frac{1-\epsilon_{\sigma(i+1)}}{2}}(x_{\sigma(i+1)}),\xi_{\sigma(i)}^{\frac{1-\epsilon_{\sigma(i)}}{2}}(y_{\sigma(i)}) \right)\right)\\
        =&\sum_{\epsilon_1=1,\epsilon_2,\ldots,\epsilon_k\in\{\pm1\}}\sum_{\sigma\in S_k'}\left( \prod_{i=1}^k\epsilon_{\sigma(i)} \right)\sum_{j=1}^k(-1)^j\left( \prod_{i=1}^{j-1}g\left( \xi_{\sigma(i+1)}^{\frac{1-\epsilon_{\sigma(i+1)}}{2}}(x_{\sigma(i+1)}),\xi_{\sigma(i)}^{\frac{1-\epsilon_{\sigma(i)}}{2}}(y_{\sigma(i)}) \right) \right)\cdot\notag\\
        &\cdot\left( 2t\left( \xi_{\sigma(j)}^{\frac{1-\epsilon_{\sigma(j)}}{2}}(y_{\sigma(j)}) \right)-1 \right)\left( 2t\left( \xi_{\sigma(j+1)}^{\frac{1-\epsilon_{\sigma(j+1)}}{2}}(x_{\sigma(j+1)}) \right)-1 \right)\cdot\notag\\
        &\cdot\left( \prod_{i=j+1}^kg\left( \xi_{\sigma(i)}^{\frac{1-\epsilon_{\sigma(i)}}{2}}(y_{\sigma(i)}),\xi_{\sigma(i+1)}^{\frac{1-\epsilon_{\sigma(i+1)}}{2}}(x_{\sigma(i+1)}) \right) \right)\\
        =&\sum_{j=1}^k\left( \sum_{\epsilon_1=1,\epsilon_2,\ldots,\epsilon_k\in\{\pm1\}}\sum_{\sigma\in S_k'}\left( \prod_{i=1}^k\epsilon_{\sigma(i)} \right)(-1)^j\left( \prod_{i=1}^{j-1}g\left( \xi_{\sigma(i+1)}^{\frac{1-\epsilon_{\sigma(i+1)}}{2}}(x_{\sigma(i+1)}),\xi_{\sigma(i)}^{\frac{1-\epsilon_{\sigma(i)}}{2}}(y_{\sigma(i)}) \right) \right)\cdot \right.\notag\\
        &\cdot\left( 2t\left( \xi_{\sigma(j)}^{\frac{1-\epsilon_{\sigma(j)}}{2}}(y_{\sigma(j)}) \right)-1 \right)\left( 2t\left( \xi_{\sigma(j+1)}^{\frac{1-\epsilon_{\sigma(j+1)}}{2}}(x_{\sigma(j+1)}) \right)-1 \right)\cdot\notag\\
        &\left. \cdot\left( \prod_{i=j+1}^kg\left( \xi_{\sigma(i)}^{\frac{1-\epsilon_{\sigma(i)}}{2}}(y_{\sigma(i)}),\xi_{\sigma(i+1)}^{\frac{1-\epsilon_{\sigma(i+1)}}{2}}(x_{\sigma(i+1)}) \right) \right) \right)\\
        =&:\sum_{j=1}^kZ_j(x_1,y_1,\ldots,x_k,y_k).
    \end{align}
\end{subequations}
Here we use the following equality:
\begin{equation}\label{split}
    \begin{aligned}
        &g(a_1,b_1)\cdot\cdots\cdot g(a_k,b_k)+(-1)^{k-1}g(b_1,a_1)\cdot\cdots\cdot g(b_k,a_k)\\
        =&g(a_1,b_1)g(a_2,b_2)\cdot\cdots\cdot g(a_k,b_k)+g(b_1,a_1)g(a_2,b_2)\cdot\cdots\cdot g(a_k,b_k)\\
        -&g(b_1,a_1)g(a_2,b_2)\cdot\cdots\cdot g(a_k,b_k)+(-1)^{k-1}g(b_1,a_1)\cdot\cdots\cdot g(b_k,a_k)\\
        =&-(2t(a_1)-1)(2t(b_1)-1)g(a_2,b_2)\cdot\cdots\cdot g(a_k,b_k)\\
        &-g(b_1,a_1)g(a_2,b_2)\cdot\cdots\cdot g(a_k,b_k)+(-1)^{k-1}g(b_1,a_1)\cdot\cdots\cdot g(b_k,a_k)\\
        =&-(2t(a_1)-1)(2t(b_1)-1)g(a_2,b_2)\cdot\cdots\cdot g(a_k,b_k)\\
        &-g(b_1,a_1)g(a_2,b_2)\cdot\cdots\cdot g(a_k,b_k)-g(b_1,a_1)g(b_2,a_2)g(a_3,b_3)\cdot\cdots\cdot g(a_k,b_k)\\
        &+g(b_1,a_1)g(b_2,a_2)g(a_3,b_3)\cdot\cdots\cdot g(a_k,b_k)+(-1)^{k-1}g(b_1,a_1)\cdot\cdots\cdot g(b_k,a_k)\\
        =&-(2t(a_1)-1)(2t(b_1)-1)g(a_2,b_2)\cdot\cdots\cdot g(a_k,b_k)\\
        &-g(b_1,a_1)(2t(a_2)-1)(2t(b_2)-1)g(a_3,b_3)\cdot\cdots\cdot g(a_k,b_k)\\
        &+g(b_1,a_1)g(b_2,a_2)g(a_3,b_3)\cdot\cdots\cdot g(a_k,b_k)+(-1)^{k-1}g(b_1,a_1)\cdot\cdots\cdot g(b_k,a_k)\\
        =&\cdots=\sum_{j=1}^k(-1)^j\left( \prod_{i=1}^{j-1}g(b_i,a_i) \right)(2t(a_j)-1)(2t(b_j)-1)\left( \prod_{i=j+1}^kg(a_i,b_i) \right).
    \end{aligned}
\end{equation}
We want to show that for all \(j=1,\ldots,k\),
\begin{align*}
    0=&\sum_{l=2}^kZ_j(x_l,y_1,x_2,y_2,\ldots,\hat{x}_l,\hat{y}_l,\ldots,x_k,y_k)(2t(y_l)-1)\\&-Z_j(y_l,y_1,x_2,y_2,\ldots,\hat{x}_l,\hat{y}_l,\ldots,x_k,y_k)(2t(x_l)-1),
\end{align*}
Since
\begin{align*}
    &Z_j(x_1,y_1,\ldots,x_k,y_k)\\
    =&\sum_{\epsilon_1=1,\epsilon_2,\ldots,\epsilon_k\in\{\pm1\}}\sum_{\sigma\in S_k'}\prod_{i=1}^k\epsilon_{\sigma(i)}(-1)^j\left( \prod_{i=1}^{j-1}g\left( \xi_{\sigma(i+1)}^{\frac{1-\epsilon_{\sigma(i+1)}}{2}}(x_{\sigma(i+1)}),\xi_{\sigma(i)}^{\frac{1-\epsilon_{\sigma(i)}}{2}}(y_{\sigma(i)}) \right) \right)\cdot\\
    &\cdot\left( 2t\left( \xi_{\sigma(j)}^{\frac{1-\epsilon_{\sigma(j)}}{2}}(y_{\sigma(j)}) \right)-1 \right)\left( 2t\left( \xi_{\sigma(j+1)}^{\frac{1-\epsilon_{\sigma(j+1)}}{2}}(x_{\sigma(j+1)}) \right)-1 \right)\cdot\\&\cdot\left( \prod_{i=j+1}^kg\left( \xi_{\sigma(i)}^{\frac{1-\epsilon_{\sigma(i)}}{2}}(y_{\sigma(i)}),\xi_{\sigma(i+1)}^{\frac{1-\epsilon_{\sigma(i+1)}}{2}}(x_{\sigma(i+1)}) \right) \right)\\
    =&(-1)^j\sum_{\sigma\in S_j'}\sum_{\epsilon_1=1,\epsilon_{\sigma(2)},\ldots,\epsilon_{\sigma(j)}\in\{\pm1\}}\left( \prod_{i=2}^j\epsilon_{\sigma(i)} \right)\left( \prod_{i=1}^{j-1}g\left( \xi_{\sigma(i+1)}^{\frac{1-\epsilon_{\sigma(i+1)}}{2}}(x_{\sigma(i+1)}),\xi_{\sigma(i)}^{\frac{1-\epsilon_{\sigma(i)}}{2}}(y_{\sigma(i)}) \right) \right)\cdot\\
    &\cdot\left( 2t\left( \xi_{\sigma(j)}^{\frac{1-\epsilon_{\sigma(j)}}{2}}(y_{\sigma(j)}) \right)-1 \right)\cdot\left( \sum_{\epsilon_1=1,\epsilon_{i_{j+1}(\sigma)},\ldots,\epsilon_{i_k(\sigma)}\in\{\pm1\}}\sum_{\eta_\sigma}\left( \prod_{m=j+1}^k\epsilon_{i_m}(\sigma) \right) \right.\\
    &\left. \left( 2t\left( \xi_{\eta_\sigma(i_{j+1})}^{\frac{1-\epsilon_{\eta_\sigma(i_{j+1})}}{2}}(x_{\eta_\sigma(i_{j+1})}) \right)-1 \right)\prod_{m=j+1}^kg\left( \xi_{\eta_\sigma(i_{m})}^{\frac{1-\epsilon_{\eta_\sigma(i_{m})}}{2}}(y_{\eta_\sigma(i_{m})}),\xi_{\eta_\sigma(i_{m+1})}^{\frac{1-\epsilon_{\eta_\sigma(i_{m+1})}}{2}}(x_{\eta_\sigma(i_{m+1})}) \right) \right)\\
    =&(-1)^j\sum_{\substack{\{i_2,\ldots,i_j\}\cup\{i_{j+1},\ldots,i_k\}\\=\{2,\ldots,k\}}}\left( \sum_{\epsilon_1=1,\epsilon_{i_2},\ldots,\epsilon_{i_j}\in\{\pm1\}}\sum_\mu\left( \prod_{m=2}^j\epsilon_{i_m} \right) \right.\\
    &\left.\left( \prod_{m=1}^{j-1}g\left( \xi_{\mu(i_{m+1})}^{\frac{1-\epsilon_{\mu(i_{m+1})}}{2}}(x_{\mu(i_{m+1})}),\xi_{\mu(i_{m})}^{\frac{1-\epsilon_{\mu(i_{m})}}{2}}(y_{\mu(i_{m})}) \right) \right)\left( 2t\left( \xi_{\mu(i_{j})}^{\frac{1-\epsilon_{\mu(i_{j})}}{2}}(y_{\mu(i_{j})}) \right)-1 \right)\right)\\
    &\cdot\left( \sum_{\epsilon_1=1,\epsilon_{i_{j+1}},\ldots,\epsilon_{i_k}\in\{\pm1\}}\sum_{\eta}\left( \prod_{m=j+1}^k\epsilon_{i_m} \right)\left( 2t\left( \xi_{\eta(i_{j+1})}^{\frac{1-\epsilon_{\eta(i_{j+1})}}{2}}(x_{\eta(i_{j+1})}) \right)-1 \right) \right.\\
    &\left. \prod_{m=j+1}^kg\left( \xi_{\eta(i_{m})}^{\frac{1-\epsilon_{\eta(i_{m})}}{2}}(y_{\eta(i_{m})}),\xi_{\eta(i_{m+1})}^{\frac{1-\epsilon_{\eta(i_{m+1})}}{2}}(x_{\eta(i_{m+1})}) \right) \right)\\
    =&:(-1)^j\sum_{\substack{\{i_2,\ldots,i_j\}\cup\{i_{j+1},\ldots,i_k\}\\=\{2,\ldots,k\}}}H(y_1,x_{i_2},y_{i_2},\ldots,x_{i_j},y_{i_j})\mathcal{G}(x_{i_{j+1}},y_{i_{j+1}},\ldots,x_{i_m},y_{i_m},x_1).
\end{align*}
where \(\eta_\sigma\) is a permutation of
\[\{i_{j+1}(\sigma),\ldots,i_k(\sigma)\}=\{2,\ldots,n\}-\{\sigma(2),\ldots,\sigma(j)\},\]
and \(\mu\) is a permutaion of \(\{i_2,\ldots,i_j\}\) and \(\eta\) is a permutaion of \(\{i_{j+1},\ldots,i_k\}\). Let
\begin{equation}\label{G}
    \begin{aligned}
        G(x_1,y_1,\ldots,x_k,y_k):=&\sum_{\epsilon_1,\ldots,\epsilon_k\in\{\pm1\}}\sum_{\sigma\in S_k'}\prod_{i=1}^k\epsilon_i\left( 2t\left( \xi_{\sigma(2)}^{\frac{1-\epsilon_{\sigma(2)}}{2}}(x_{\sigma(2)}) \right)-1 \right)\cdot\\
        &\cdot\left( \prod_{i=2}^{k}g\left( \xi_{\sigma(i)}^{\frac{1-\epsilon_{\sigma(i)}}{2}}(y_{\sigma(i)}),\xi_{\sigma(i+1)}^{\frac{1-\epsilon_{\sigma(i+1)}}{2}}(x_{\sigma(i+1)}) \right) \right)\left( 2t\left( \xi_1^{\frac{1-\epsilon_1}{2}}(y_1) \right)-1 \right).
    \end{aligned}
\end{equation}
Then
\begin{subequations}\label{ZHG}
    \begin{align}
        &\sum_{l=2}^k\left( Z_j(x_l,y_1,x_2,y_2,\ldots,\hat{x}_l,\hat{y}_l,\ldots,x_k,y_k)(2t(y_l)-1)\right.\\
        &\left.-Z_j(y_l,y_1,x_2,y_2,\ldots,\hat{x}_l,\hat{y}_l,\ldots,x_k,y_k)(2t(x_l)-1) \right)\\
        =&(-1)^j\sum_{l=2}^k\sum_{\substack{\{i_2,\ldots,i_j\}\cup\{i_{j+1},\ldots,i_{k-1}\}\\=\{2,\ldots,\hat{l},\ldots,k\}}}H(y_1,x_{i_2},y_{i_2},\ldots,x_{i_j},y_{i_j})\\
        &\left( \mathcal{G}(x_{i_{j+1}},y_{i_{j+1}},\ldots,x_{i_{k-1}},y_{i_{k-1}},x_l)(2t(y_l)-1)-\mathcal{G}(x_{i_{j+1}},y_{i_{j+1}},\ldots,x_{i_{k-1}},y_{i_{k-1}},y_l)(2t(x_l)-1) \right)\\
        =&(-1)^j\sum_{l=2}^k\sum_{\substack{\{i_2,\ldots,i_j\}\cup\{i_{j+1},\ldots,i_{k-1}\}\\=\{2,\ldots,\hat{l},\ldots,k\}}}H(y_1,x_{i_2},y_{i_2},\ldots,x_{i_j},y_{i_j})G(x_l,y_l,x_{i_{j+1}},y_{i_{j+1}},\ldots,x_{i_{k-1}},y_{i_{k-1}})\\
        =&(-1)^j\sum_{\substack{\{i_2,\ldots,i_j\}\cup\{i_{j+1},\\\ldots, i_{k}\}=\{2,\ldots,k\}}}H(y_1,x_{i_2},y_{i_2},\ldots,x_{i_j},y_{i_j})\\&
        \left( \sum_{m=j+1}^kG(x_{i_m},y_{i_m},x_{i_{j+1}},y_{i_{j+1}},\ldots,\hat{x}_{i_m},\hat{y}_{i_m},\ldots,x_{i_k},y_{i_k}) \right)\label{HG}
    \end{align}
\end{subequations}
\eqref{Fn0} tells us that
\begin{align*}
    &\sum_{l=2}^nG(x_l,y_l,x_2,y_2,\ldots,\hat{x}_l,\hat{y}_l,\ldots,x_n,y_n)\\
    =&\sum_{l=2}^n\sum_{\epsilon_2,\ldots,\epsilon_n\in\{\pm1\}}\sum_{\substack{\sigma\in S_n\\\sigma(1)=1,\sigma(l)=l}}\prod_{i=2}^n\epsilon_i\cdot\left( 2t\left( \xi_{\sigma(2)}^{\frac{1-\epsilon_{\sigma(2)}}{2}}(x_{\sigma(2)}) \right)-1 \right)\\&\cdot\left( \prod_{i=2}^{l-2}g\left( \xi_{\sigma(i)}^{\frac{1-\epsilon_{\sigma(i)}}{2}}(y_{\sigma(i)}),\xi_{\sigma(i+1)}^{\frac{1-\epsilon_{\sigma(i+1)}}{2}}(x_{\sigma(i+1)}) \right) \right)g\left( \xi_{\sigma(l-1)}^{\frac{1-\epsilon_{\sigma(l-1)}}{2}}(y_{\sigma(l-1)}),\xi_{\sigma(l+1)}^{\frac{1-\epsilon_{\sigma(l+1)}}{2}}(x_{\sigma(l+1)}) \right)\\
    &\left( \prod_{i=l+1}^{n-1}g\left( \xi_{\sigma(i)}^{\frac{1-\epsilon_{\sigma(i)}}{2}}(y_{\sigma(i)}),\xi_{\sigma(i+1)}^{\frac{1-\epsilon_{\sigma(i+1)}}{2}}(x_{\sigma(i+1)}) \right) \right)g\left( \xi_{\sigma(n)}^{\frac{1-\epsilon_{\sigma(n)}}{2}}(y_{\sigma(n)}),\xi_{l}^{\frac{1-\epsilon_{l}}{2}}(x_{l}) \right)\\
    &\left( 2t\left( \xi_{l}^{\frac{1-\epsilon_{l}}{2}}(y_{l}) \right)-1 \right)\\
    =&\sum_{\substack{(l,h)\in\{2,\ldots,n\}^2\\ l\neq h}}\sum_{\epsilon_2,\ldots,\epsilon_n\in\{\pm1\}}\sum_{\substack{\sigma\in S_n,\sigma(1)=1,\\\sigma(2)=h,\sigma(l)=l}}\prod_{i=2}^n\epsilon_i\cdot\left( 2t\left( \xi_{h}^{\frac{1-\epsilon_{h}}{2}}(x_{h}) \right)-1 \right)\\&\cdot\left( \prod_{i=2}^{l-2}g\left( \xi_{\sigma(i)}^{\frac{1-\epsilon_{\sigma(i)}}{2}}(y_{\sigma(i)}),\xi_{\sigma(i+1)}^{\frac{1-\epsilon_{\sigma(i+1)}}{2}}(x_{\sigma(i+1)}) \right) \right)g\left( \xi_{\sigma(l-1)}^{\frac{1-\epsilon_{\sigma(l-1)}}{2}}(y_{\sigma(l-1)}),\xi_{\sigma(l+1)}^{\frac{1-\epsilon_{\sigma(l+1)}}{2}}(x_{\sigma(l+1)}) \right)\\
    &\left( \prod_{i=l+1}^{n-1}g\left( \xi_{\sigma(i)}^{\frac{1-\epsilon_{\sigma(i)}}{2}}(y_{\sigma(i)}),\xi_{\sigma(i+1)}^{\frac{1-\epsilon_{\sigma(i+1)}}{2}}(x_{\sigma(i+1)}) \right) \right)g\left( \xi_{\sigma(n)}^{\frac{1-\epsilon_{\sigma(n)}}{2}}(y_{\sigma(n)}),\xi_{l}^{\frac{1-\epsilon_{l}}{2}}(x_{l}) \right)\\
    &\left( 2t\left( \xi_{l}^{\frac{1-\epsilon_{l}}{2}}(y_{l}) \right)-1 \right)\\
    =&\left(  \sum_{2\leq l<h\leq n}\left((2t(x_h)-1)\tilde{Z}(x_l,y_h,x_2,y_2,\ldots,\hat{x}_l,\hat{y}_l,\ldots,\hat{x}_h,\hat{y}_h,\ldots,x_n,y_n)(2t(y_l)-1) \right. \right.\\
    &+(2t(y_h)-1)\tilde{Z}(y_l,x_h,x_2,y_2,\ldots,\hat{x}_l,\hat{y}_l,\ldots,\hat{x}_h,\hat{y}_h,\ldots,x_n,y_n)(2t(x_l)-1)\\
    &-(2t(x_h)-1)\tilde{Z}(y_l,y_h,x_2,y_2,\ldots,\hat{x}_l,\hat{y}_l,\ldots,\hat{x}_h,\hat{y}_h,\ldots,x_n,y_n)(2t(x_l)-1)\\
    &\left. -(2t(y_h)-1)\tilde{Z}(x_l,x_h,x_2,y_2,\ldots,\hat{x}_l,\hat{y}_l,\ldots,\hat{x}_h,\hat{y}_h,\ldots,x_n,y_n)(2t(y_l)-1) \right)\\
    &+  \sum_{2\leq h<l\leq n}\left((2t(x_h)-1)\tilde{Z}(x_l,y_h,x_2,y_2,\ldots,\hat{x}_h,\hat{y}_h,\ldots,\hat{x}_l,\hat{y}_l,\ldots,x_k,y_k)(2t(y_l)-1) \right.\\
    &+(2t(y_h)-1)\tilde{Z}(y_l,x_h,x_2,y_2,\ldots,\hat{x}_h,\hat{y}_h,\ldots,\hat{x}_l,\hat{y}_l,\ldots,x_n,y_n)(2t(x_l)-1)\\
    &-(2t(x_h)-1)\tilde{Z}(y_l,y_h,x_2,y_2,\ldots,\hat{x}_h,\hat{y}_h,\ldots,\hat{x}_l,\hat{y}_l,\ldots,x_n,y_n)(2t(x_l)-1)\\
    &\left.\left. -(2t(y_h)-1)\tilde{Z}(x_l,x_h,x_2,y_2,\ldots,\hat{x}_h,\hat{y}_h,\ldots,\hat{x}_l,\hat{y}_l,\ldots,x_n,y_n)(2t(y_l)-1) \right)\right)\\
    =&\frac{1}{2}
    \sum_{l=2}^n\left( (2t(x_l)-1)F(y_l,x_2,y_2,\ldots,\hat{x}_l,\hat{y}_l,\ldots,x_n,y_n)\right.\\
    &\left.+(-1)^{n-2}(2t(y_l)-1)F(x_l,y_2,x_2,\ldots,\hat{y}_l,\hat{x}_l,\ldots,y_n,x_n) \right)\\
    =&0.
\end{align*}
From this result, we know that \eqref{HG} is 0. Hence
\begin{align*}
    &F(y_1,x_2,y_2,\ldots,x_k,y_k)\\
    =&\sum_{j=1}^k\sum_{l=2}^k\left( Z_j(x_l,y_1,x_2,y_2,\ldots,\hat{x}_l,\hat{y}_l,\ldots,x_k,y_k)(2t(y_l)-1)\right.\\&\left.-Z_j(y_l,y_1,x_2,y_2,\ldots,\hat{x}_l,\hat{y}_l,\ldots,x_k,y_k)(2t(x_l)-1) \right)\\
    =&0.\qedhere
\end{align*}

\addcontentsline{toc}{section}{References}
\newcommand{\etalchar}[1]{$^{#1}$}

\end{document}